\documentclass[reprint, aps, pre, 10pt, twocolumn, amsmath, amssymb]{revtex4-2}
\usepackage{amsfonts}
\usepackage[T1]{fontenc}
\usepackage{multirow}
\usepackage{graphicx}
\usepackage{esint}
\usepackage[percent]{overpic}
\usepackage{xcolor}
\usepackage[hidelinks, bookmarks=false]{hyperref}

\usepackage[caption=false]{subfig}
\usepackage{xcolor}
\hypersetup{
    colorlinks,
    linkcolor={blue!80!black},
    citecolor={blue!80!black},
    urlcolor={blue!80!black}
}

\usepackage{booktabs} 
\usepackage{graphicx}
\usepackage{enumerate}
\usepackage{comment}
\usepackage{color}

\usepackage{tikz}
\usetikzlibrary{arrows.meta,calc,positioning,shadows.blur,backgrounds}

\definecolor{cX}    {RGB}{ 21, 86,180}   
\definecolor{cY}    {RGB}{ 16,124, 78}   
\definecolor{cXY}   {RGB}{124, 32,146}   
\definecolor{cYX}   {RGB}{214,126,  6}   
\definecolor{cSite} {RGB}{ 24, 27, 35}   
\definecolor{cHub}  {RGB}{214, 40, 40}   
\definecolor{cGrid} {RGB}{198,204,214}   
\definecolor{cPanel}{RGB}{250,251,253}   
\definecolor{cEdge} {RGB}{220,225,233}   
\definecolor{cInk}  {RGB}{ 22, 24, 29}   
\definecolor{cMute} {RGB}{ 98,106,120}   

\begin{document}

\title{Odd diffusion and power-law correlations in chiral mass-transport processes}

\author{Koushik Das}
\email{koushik.das@bose.res.in}
\author{Animesh Hazra}
\author{Punyabrata Pradhan}
\affiliation{Department of Physics of Complex Systems, S. N. Bose National Centre for Basic Sciences, Block-JD, Sector-III, Salt Lake, Kolkata, 700106, India.}

\begin{abstract}

We study mass-conserving Markov jump processes on a square lattice, where masses hop with a preferred rotational sense, thus {\it breaking both time-reversal and mirror symmetries.} 
We consider closed systems with both periodic and open (reflecting) boundaries, the latter supporting a steady-state edge current. We show that odd diffusion, arising from the chiral transport, generically provides a mechanism for the emergence of scale-invariant two-point density correlations in nonequilibrium steady states even in the {\it presence of lattice rotation symmetry}—a mechanism that is qualitatively distinct from the well-known mechanism of anisotropic hopping.  
We exactly calculate the steady-state equal-time density-density correlations, which exhibit an algebraic decay, $C(\mathbf r)\sim |\mathbf r|^{-4}$ for $|\mathbf r|\gg1$. This algebraic behavior results from the interplay between the {\it off-diagonal} components of the diffusion and mobility tensors, demonstrating that chiral transport alone can generate power-law correlations in isotropic driven systems.
Remarkably, the structure factor in the zero-wavenumber limit and the amplitude of the power laws depend nontrivially on chirality. While increasing chirality initially suppresses large-scale density fluctuations, the fluctuations beyond a threshold odd-diffusion strength vary nonmonotonically with chirality and  develop a cusp singularity.

\end{abstract}

\maketitle

\section{Introduction}
\label{Introduction}

Transport phenomena in systems driven far from equilibrium continue to reveal a remarkable variety of collective behaviors with no counterpart in equilibrium systems. Examples range from active matter \cite{Marchetti2013Jul} and driven colloidal suspensions \cite{Ginot2025Nov} to biological transport \cite{Doyle1998Apr}, electronic systems \cite{Dutta1981Jul}, and complex fluids \cite{Leal2007Jun, Gelbart1996Aug}. In recent years, particular attention has been devoted to chiral systems, which break both time-reversal and mirror (parity) symmetries and consequently exhibit transport properties that are qualitatively different from those of their achiral counterparts \cite{Fruchart2023Mar}. Chiral transport has been investigated in a broad range of physical settings, including active rotors \cite{Nguyen2014Feb}, glassy dynamics \cite{Guo2025Jan} and circular motion \cite{Kummel2013May} of active colloids, and various topological materials \cite{Hasan2010Nov}. A hallmark of these systems is the emergence of rotational currents and transverse transport.

The notion of odd transport is well known, and the associated transport coefficients have long been studied in condensed-matter physics \cite{Avron1998Aug}, particularly in the contexts of Hall transport and odd viscosity \cite{Avron1995Jul}, where broken time-reversal symmetry gives rise to nondissipative transport \cite{Miyake2013Oct, Hugel2014Feb}. By contrast, much less is known about the role of odd transport in determining fluctuation properties of nonequilibrium many-body systems \cite{Lei2019Jan, Banerjee2022Apr}.
A fundamental question is whether odd diffusion alone can qualitatively alter the large-scale structure of steady-state density correlations. Here, we address this question in isotropic systems with mass-conserving dynamics (i.e., closed systems) and with no net bulk current, so that the anisotropy that conventionally provides a mechanism for generic algebraic correlations \cite{Garrido1990Aug, Grinstein1990Apr} is absent. We demonstrate that odd diffusion provides an alternative, robust mechanism for generating power-law density correlations in isotropic nonequilibrium systems.

The recent surge of interest in odd transport has been driven largely by the discovery of striking and often counterintuitive collective phenomena in active matter \cite{Liebchen2017Aug, Liebchen2022Sep}. These include odd viscosity \cite{Eren2026Aug},  topological edge currents \cite{vanZuiden2016Nov, Soni2019Nov}, pattern formation and phase transitions \cite{Levis2019Jul, Caporusso2024Apr, Caprini2025Apr, Wang2024SepNJP, Dutta2026Sep}, and unconventional relaxation dynamics \cite{VegaReyes2022Oct, Kalz2022Aug, Kalz2024Jan}, all arising from odd interactions. 
Experiments on chiral active matter have further revealed persistent edge currents and boundary-localized transport \cite{Chen2024Mar, Yashunsky2022Nov}, highlighting the intriguing consequences of broken parity and time-reversal symmetries in chiral systems. Motivated by these developments, considerable effort has been devoted to developing a general theoretical framework for odd transport; see, e.g., Refs. \cite{Lowen2016Nov, Shankar2022Jun}.

Despite recent progress, a rigorous theoretical understanding of the role of chirality and many-body interactions in determining fluctuation properties remains lacking.
Existing exact studies have largely focused on noninteracting systems, such as paradigmatic models of a chiral random walker (or, a chiral Brownian particle) \cite{Larralde1997Nov, Hargus2021Oct, Sevilla2016Dec, Caprini2019Mar}. While these models provide valuable insight, elucidating the consequences of broken symmetries, they cannot capture the collective fluctuations and correlations that emerge from many-body interactions. On the other hand, introducing interactions generally renders the problem analytically intractable, making exact results exceedingly rare. There is therefore a clear need for exactly solvable interacting models that can elucidate the interplay between chirality, many-body interactions, and fluctuations in these systems. 
To this end, we introduce a class of Markov jump processes with nontrivial spatial structure that nevertheless remain exactly solvable. This exact solvability allows us to determine the transport coefficients governing the large-scale behavior directly from the underlying microscopic dynamics, thereby providing a microscopic understanding of the emergence of large-scale fluctuations in chiral nonequilibrium systems within a minimal setting.

Another particularly important question, recently explored in the literature, concerns the existence of fluctuation-dissipation relations (FDRs) in chiral systems driven away from equilibrium \cite{Hargus2021Oct, Kalz2022Aug, Hargus2025Mar}. Near equilibrium, the celebrated Green-Kubo relations establish a direct connection between transport coefficients and equilibrium current fluctuations \cite{Chun2021Dec, Cengio2019Dec}. More generally, FDRs provide one of the cornerstones of statistical mechanics by linking spontaneous fluctuations to linear response \cite{Marconi2008March}. Whether analogous relations exist in driven systems that explicitly violate time-reversal symmetry remains a fundamental issue. Although several generalized FDRs have been proposed for nonequilibrium steady states \cite{Speck2006May, Baiesi2009Jul}, an exact microscopic derivation of transport coefficients, together with a systematic connection between these coefficients and density fluctuations that leads to such relations, remains largely elusive in interacting nonequilibrium systems.

In this work, we address these issues by introducing minimal models of interacting chiral systems—a family of generalized Kipnis–Marchioro–Presutti (KMP) models that remain amenable to exact analytical treatment.
The original KMP model has long served as a paradigmatic framework for studying transport and fluctuation phenomena \cite{Bertini2015Jun}. Here, we generalize the model by constructing mass-conserving Markov jump processes in which mass transfer occurs with a preferred rotational sense, either clockwise or counterclockwise. The resulting bulk dynamics remain diffusive, but explicitly break both time-reversal and mirror (parity) symmetries while preserving mass conservation. Remarkably, despite exhibiting nontrivial many-body correlations, these systems remain exactly solvable, making them an ideal testing ground for studying nonequilibrium fluctuations in interacting chiral systems.

We study the relaxation and fluctuation properties of these chiral mass-transport processes on a two-dimensional square lattice. At large spatio-temporal scales, density fluctuations evolve diffusively, characterized by the bulk-diffusion and Onsager (mobility) tensors. However, the broken symmetries give rise to fundamentally new fluctuation properties. We show that, {\it even for isotropic mass-transfer rules}, chiral hopping generically gives rise to algebraic equal-time density-density correlations in the steady state of mass-conserving systems; notably, this particular mechanism of generating the algebraic correlations is qualitatively different from the well-known mechanism of anisotropic hopping \cite{Grinstein1990Apr, Garrido1990Aug}. To characterize these correlations, we derive exact expressions for the two-point spatial density correlation function in terms of the diffusion and mobility tensors. The resulting theoretical framework also leads to a nonequilibrium analog of the celebrated Green-Kubo relation, thus establishing a direct connection between transport coefficients and steady-state fluctuations in these chiral systems.

Our principal finding is that the steady-state density correlations exhibit algebraic decay at large distances. Specifically, we show that the correlation function has the asymptotic form
\begin{align}
C({\bf r}) \simeq A(\theta) |{\bf r}|^{-\eta},
\end{align}
where, for an isotropic system on a two-dimensional square lattice, we find $\eta=4$, indicating the emergence of scale-invariant fluctuations. The prefactor $A(\theta)$ captures the angular dependence of the correlations. Indeed, the interplay between the {\it off-diagonal} elements of the diffusion and mobility tensors plays a crucial role in generating this algebraic decay, demonstrating that chiral transport alone can produce power-law correlations even in an otherwise {\it isotropic} system.

Furthermore, we find that the odd-diffusion coefficient $D^{(\rm odd)}$ plays a dual role in determining density fluctuations. At small strength, it suppresses long-wavelength fluctuations (or, equivalently, the zero-wave-vector structure factor). As the maximal admissible value of odd diffusion, {$D_c^{(\rm odd)}$, is approached, however, the mobility and ``compressibility'' start increasing and then develop a cusp-like singularity, implying an increasingly strong response to a small change in the odd-diffusion coefficient. Thus, chirality initially stabilizes the system by promoting mixing and reducing density inhomogeneities, but, beyond an optimal value, it drives the system toward a singular regime where the fluctuations become exceptionally sensitive to odd transport.

The rest of the paper is organized as follows. In Sec.~\ref{sec:model}, we introduce a broad class of chiral mass-transport processes where we incorporate a rotational sense to the microscopic hopping dynamics. In Section~\ref{sec:density_evol}, we develop a microscopic theory and derive the diffusion tensor; subsequently, in Sec.~\ref{Curr-fluc}, we derive the mobility tensor. In Sec.~\ref{subsec:dens_corl}, we then obtain exact expressions for the steady-state two-point density correlation functions and analyze their asymptotic behavior. In Sec.~\ref{subsec:FDR}, we derive a generalized Einstein relation connecting the transport coefficients and zero-wavenumber structure factor. Finally, in Sec.~\ref{sec:summary}, we summarize the results with some concluding remarks.

\section{Model}\label{sec:model}

In this section, we introduce a class of chiral many-body systems, referred to as chiral mass chipping models (MCMs), which are described by interacting Markov jump processes on a two-dimensional square lattice. These processes belong to the family of generalized Kipnis–Marchioro–Presutti (KMP) models, which have been extensively studied over the past several decades \cite{Kipnis1982Jan, Aldous1995Jun, Coppersmith1996May, Rajesh2000May, Bondyopadhyay2012Jul, Redig2017Oct}.
The lattice sites, labeled by $(x,y)$, carry a nonnegative continuous mass $m(x,y)\geq 0$. The system evolves through stochastic dynamics where the bulk mass-transfer rates are symmetric along a principal axis but explicitly break mirror (parity) symmetry [see Fig. \eqref{fig:model}].

As in the original KMP model \cite{Kipnis1982Jan}, the dynamics involve redistribution and mixing of masses among neighboring sites; that is, masses at two nearest-neighbor sites are mixed and redistributed. However, for the models considered here, masses at {\it three} neighboring sites—nearest as well as next-nearest or diagonal neighbors—are mixed and redistributed {\it simultaneously}. Notably, unlike the KMP model, the redistribution rules considered here are {\it chiral}, endowing the mass transport with a preferred handedness. This immediately results in violation of detailed balance, and consequently a non-Gibbsian steady-state measure, which, as we show later, generically lead to a scale-invariant spatial structure.


\begin{figure}[t]
\centering
\includegraphics[width = 0.95 \columnwidth]{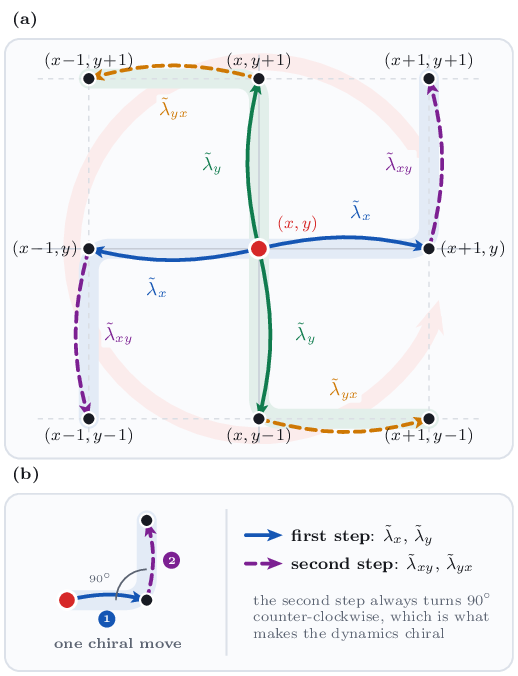}
\caption{
{\it Schematic diagram of the chiral dynamics on a square lattice.}
(a) During a microscopic time, mass transfer from the site $(x,y)$ (red solid circle)
along two orthogonal (``L''-shaped) bonds {\it simultaneously}, and always in
the {\it counter-clockwise} sense. Solid arrows correspond to the
{\it first-step} transfer, from $(x,y)$ to a nearest-neighbor site (black solid
circles) along the $x$ (blue) or $y$ (green) direction; dashed arrows
correspond to the {\it second-step} transfer, which carries the mass onward
from that neighbor to a next-nearest-neighbor site along the perpendicular
direction (purple and orange), and which introduces chirality in the dynamics.
The pale bands trace the four ``L''-shaped paths and the faint ring marks
their common sense of rotation: The four paths are $90^{\circ}$ rotations of
one another, so the dynamics is invariant under lattice rotations but not under
reflection. The microscopic parameters $\tilde\lambda_x$ ($\tilde\lambda_y$)
and $\tilde\lambda_{xy}$ ($\tilde\lambda_{yx}$) are the fractions of mass
transferred at the first and the second step, respectively. (b) A single
chiral move, magnified: The first step (badge 1) is followed by a $90^{\circ}$
left turn at the intermediate site and then the second step (badge 2).
}
\label{fig:model}
\end{figure}

This simultaneous mass transfer involving multiple bonds is \textit{necessary}, rather than merely a modeling choice. A single bond carries no intrinsic orientation and therefore dynamics constructed from independent single-bond mass transfers lead to a diagonal diffusion tensor and cannot support odd diffusion. A transverse current, perpendicular to the density gradient, requires mass transfers (or, equivalently, bond currents) along two orthogonal directions to be correlated within a single elementary event. On a square lattice, the smallest such object is an ordered pair of orthogonal bonds—an ``L''-shaped move spanning three sites—whose ordering incorporates a handedness in the problem. 

{\it Definition of the models.$-$} During a mass-transfer event, a lattice site $(x,y)$ is selected at random. The masses at the selected site, together with a randomly selected nearest neighbor and a next-nearest (diagonal) neighbor, are then mixed and redistributed according to stochastic rules that conserve the total mass of the three-site cluster. The redistribution proceeds through one of the following two moves, each of which consists of a two-step update. To make the model a generalized one, the four directions $+x$, $-x$, $+y$ and $-y$ are chosen with the probabilities $p_1, p_2, p_3, p_4 \in [0,1]$, respectively, satisfying $\sum_i p_i =1$ in the bulk.

{\it (1) Transport along $x$-direction.} (A)
Mass $ m(x,y)$ at a randomly selected site $(x,y)$ is fragmented into two fractions: A fraction $\lambda_x$ of mass is retained at the site; the remaining fraction $\tilde \lambda_x =(1-\lambda_x)$ of mass is then split such that a random fraction $\xi_1$ again returns to the site, while the remaining fraction $\tilde \xi_1=(1-\xi_1)$ is transferred to one of the nearest neighbors $(x + 1,y)$ and $(x - 1,y)$, with probability $p_1$ or $p_2$, respectively. Overall, in the first step, $m'(x,y)=(\lambda_x + \xi_1 \tilde \lambda_x) m(x,y)$ amount of mass is retained at site $(x,y)$ and the rest, $\tilde \xi_1 \tilde \lambda_x m(x,y)$, is transferred to one of its two nearest-neighbor sites.
\\
\\
(B) Then, in the second step, the chipped-off mass $\tilde \xi_1 \tilde \lambda_x m(x,y)$ mixes with the mass at a nearest neighbor site $(x\pm 1,y)$ and undergoes the following redistribution-and-mixing process: A fraction $\lambda_{xy}$ of the modified mass $m'(x \pm 1,y)=[\tilde \xi_1 \tilde \lambda_x m(x,y) + m(x\pm 1,y)]$ at the nearest-neighbor site is retained there, while the remaining fraction $\tilde \lambda_{xy}=(1-\lambda_{xy})$ of mass is further split into two random fractions, where $\xi_2$ fraction is retained and $\tilde \xi_2=(1-\xi_2)$ is transferred in $y$-direction to the next-nearest (diagonal) neighbor according to one of the following chiral (anticlockwise) transfers:
\begin{align}
(x+1,y) &\to (x+1,y+1) ~~{\rm or}~~ 
\nonumber\\
(x-1,y) &\to (x-1,y-1).
\end{align}
Therefore, in the second step, overall $m''(x \pm 1,y)=(\lambda_{xy} + \xi_2 \tilde \lambda_{xy}) m'(x \pm 1,y)$ is retained and the rest of the amount, $\tilde \xi_2 \tilde \lambda_{xy} m'(x \pm 1,y)$, is transferred in $y$-direction to a next-nearest neighbor (diagonal) site $(x\pm1,y\pm1)$, thus adding up mass at the diagonal site to $m'(x\pm1,y\pm1)=m(x\pm1,y\pm1)+\tilde \xi_2 \tilde \lambda_{xy} m'(x \pm 1,y)$. 
\\\\
Note that the above two steps (A) and (B) are executed simultaneously within an infinitesimal time interval $(t, t+dt)$. Therefore, all together, the following transition happen in the configuration space, involving redistribution of masses at three sites with the corresponding rate $p_1$ or $p_2$:
\begin{align}
    \{m(x,y), m(x \pm 1,y), m(x\pm 1, y\pm 1)\} \xrightarrow{p_1 ~{\rm or}~ p_2} \notag
    \\ \{m'(x,y), m''(x \pm 1,y), m'(x\pm 1, y\pm 1)\},
\end{align}
where the redistributed masses at time $t+dt$ are related to the initial masses at time $t$ as 
\begin{align}
    m'(x,y) &= (\lambda_x + \xi_1 \tilde \lambda_x) m(x,y),
    \\
    m''(x \pm 1,y) &= (\lambda_{xy} + \xi_2 \tilde \lambda_{xy})\notag 
    \\
    &\times [\tilde \xi_1 \tilde \lambda_x m(x,y) + m(x\pm 1,y)],
    \\
    m'(x\pm1,y\pm1) &= m(x\pm1,y\pm1)\notag
    \\
    &+\tilde \xi_2 \tilde \lambda_{xy} [\tilde \xi_1 \tilde \lambda_x m(x,y) + m(x\pm 1,y)].
\end{align}
Equivalently, we can introduce a (stochastic) mixing matrix, which relates masses $m(x,y,t+dt)$ and $m(x,y,t)$ at two infinitesimally separated times $t+dt$ (final) and $t$ (initial), respectively, as given below:
\begin{widetext}
\begin{align}
    \begin{pmatrix}
        m(x,y,t+dt) \\
        m(x\pm 1, y,t+dt)\\
        m(x\pm1, y\pm 1,t+dt)
    \end{pmatrix}
    =
    \begin{pmatrix}
       (\lambda_x + \xi_1 \tilde \lambda_x) && 0 && 0\\
       (\lambda_{xy} + \xi_2 \tilde \lambda_{xy})\tilde \xi_1 \tilde \lambda_x && (\lambda_{xy} + \xi_2 \tilde \lambda_{xy}) && 0\\
       \tilde \xi_2 \tilde \lambda_{xy}\tilde \xi_1 \tilde \lambda_{x}  && \tilde \xi_2 \tilde \lambda_{xy} && 1\\
    \end{pmatrix}
    \begin{pmatrix}
        m(x,y,t) \\
        m(x\pm 1, y,t) \\
        m(x\pm1, y\pm 1,t)
    \end{pmatrix} .
    \label{mixing-matrix}
\end{align}
\end{widetext}
{\it (2) Transport along $y$-direction.} This proceeds similar to that along $x$-direction. That is, a fraction $\lambda_y$ of mass $ m(x,y)$ is retained at site $(x,y)$, while the remaining fraction of mass, $(1-\lambda_y)m(x,y)$, is split such that a fraction $\xi_3$ returns to the site and the rest is transferred along $y$-direction to one of its two nearest neighbors $(x,y+1)$ and $(x,y-1)$ with probability $p_3$ or $p_4$, respectively. 
At the intermediate site $(x,y\pm1)$, the total mass is again redistributed: A fraction $\lambda_{yx}$ is retained, while the remaining $(1-\lambda_{yx})$ is split into a retained fraction $\xi_4$ and a transferred fraction $(1-\xi_4)$, the latter being sent according to the anticlockwise rule
\begin{align}
(x,y+1) &\to (x-1,y+1), \nonumber\\
(x,y-1) &\to (x+1,y-1).
\end{align}
One can write the above mass distribution in an infinitesimal time interval $(t, t+dt)$ through a mixing matrix similar to that given in Eq. \eqref{mixing-matrix}.

The constants $\lambda_x$, $\lambda_y$, $\lambda_{xy}$, and $\lambda_{yx}$ are various mass retention parameters, while $\xi_1$,  $\xi_2$, $\xi_3$ and $\xi_4$ are independent random variables uniformly distributed in unit interval $[0,1]$ with $n$-th moment $\mu_n = \langle \xi_i^n \rangle= 1/(n+1)$.  The second-step mixing and redistribution induces a chiral bias, thus breaking detailed balance as well as mirror symmetry.

We study these systems with both periodic and open boundary conditions, where total mass remains conserved. Under periodic boundary conditions, the dynamics is purely diffusive, and consequently the steady-state bulk current vanishes. With open (reflecting) boundaries (analogous to ``walls''), however, the system supports a nonzero steady-state edge current that flows tangentially along the boundary (wall).

\section{Density evolution}\label{sec:density_evol}

On large spatio-temporal scales, the microscopic relaxation processes due to the stochastic dynamics could be described in terms of a fluctuating hydrodynamic equation for the conserved mass density field. This description follows from local conservation laws, which relate the time evolution of coarse-grained fields to the divergence of associated currents.
For a conserved scalar density $\rho(\mathbf{r},t)$, the dynamics is governed by the continuity equation,
\begin{equation}
\frac{\partial}{\partial t} \rho(\mathbf{r},t)
= - \nabla \cdot \langle \mathbf{j}(\mathbf{r},t) \rangle
= - \nabla \cdot \left[ \langle \mathbf{j}^{(\mathrm{d})}(\mathbf{r},t) 
+ \mathbf{j}^{(\mathrm{fl})}(\mathbf{r},t) \rangle \right  ],
\label{eq:continuity}
\end{equation}
where $\mathbf{j}^{(\mathrm{d})}$ denotes the deterministic (diffusive) current and $\mathbf{j}^{(\mathrm{fl})}$ represents fluctuations, with $\langle \mathbf{j}^{(\mathrm{fl})} \rangle = 0$. The constitutive relation for the current encodes the transport properties of the system.
Beginning with a microscopic dynamical evolution of local mass, we first characterize the diffusive current component; the fluctuating current component will be characterized later in Sec. \eqref{Curr-fluc}.

\begin{figure*}
\centering

\begin{minipage}{0.32\linewidth}
\centering
\begin{overpic}[width=\linewidth]{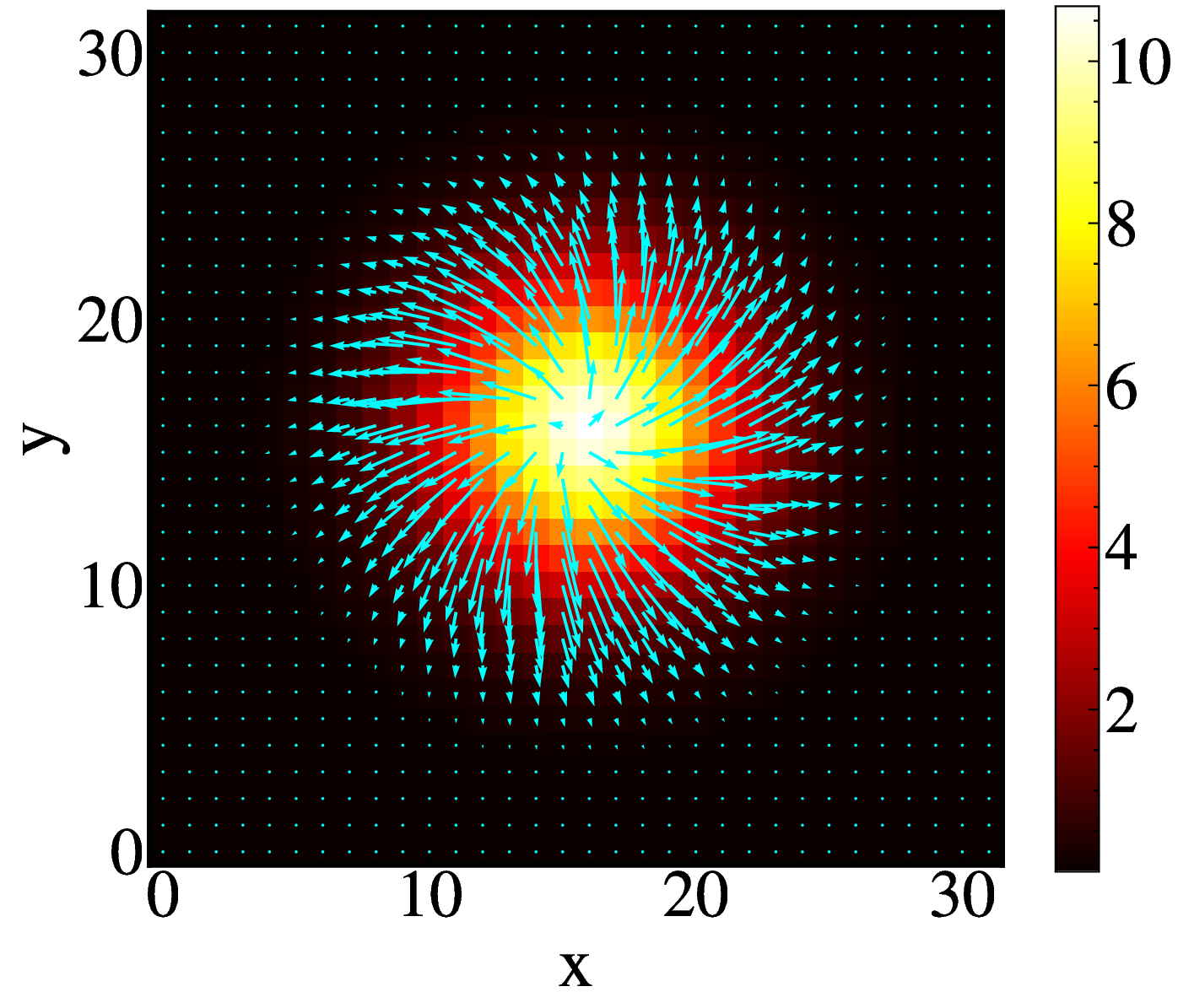}
\put(4,87){\large\bfseries (a)}
\end{overpic}
\end{minipage}
\hfill
\begin{minipage}{0.32\linewidth}
\centering
\begin{overpic}[width=\linewidth]{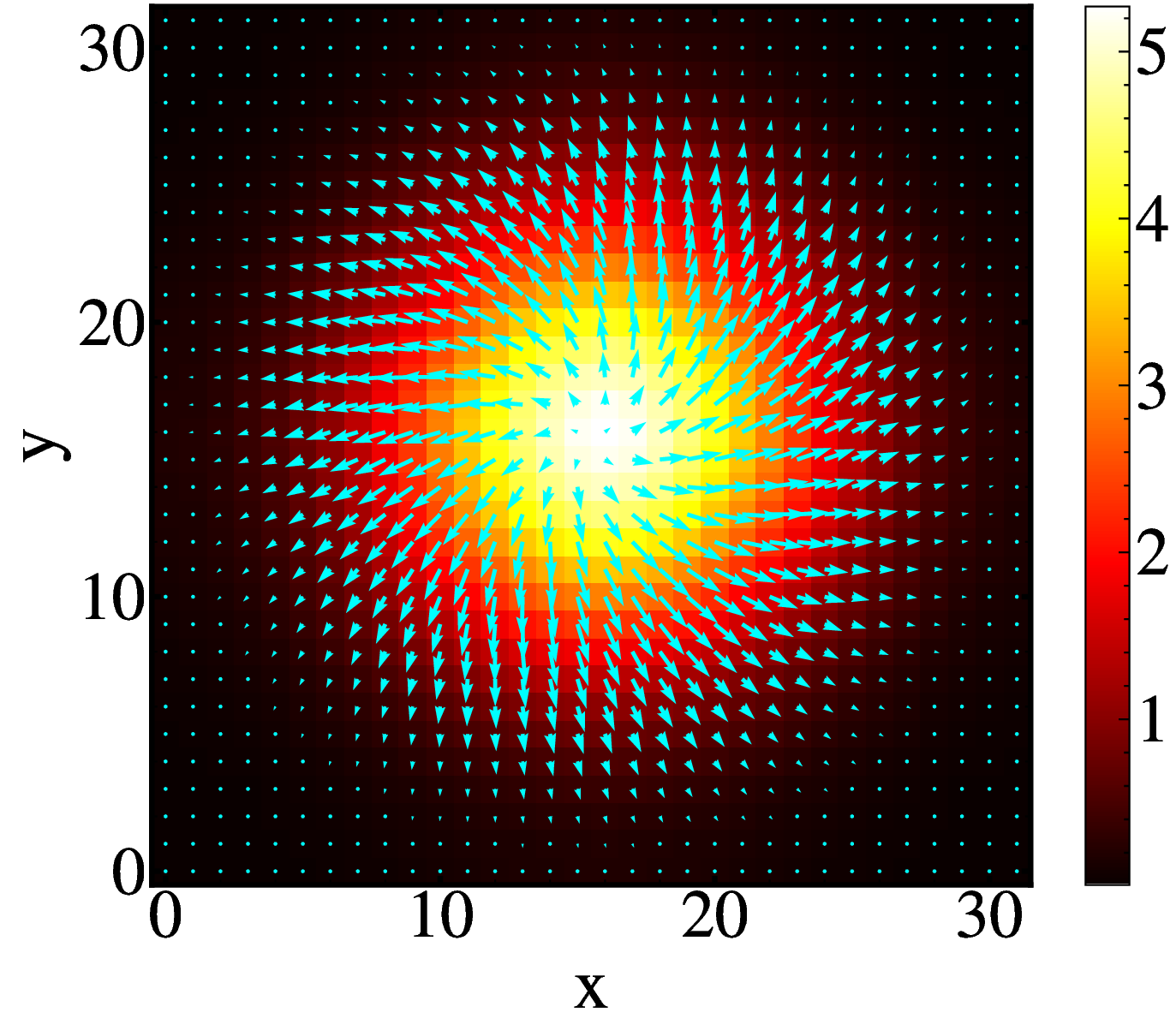}
\put(4,87){\large\bfseries (b)}
\end{overpic}
\end{minipage}
\hfill
\begin{minipage}{0.32\linewidth}
\centering
\begin{overpic}[width=\linewidth]{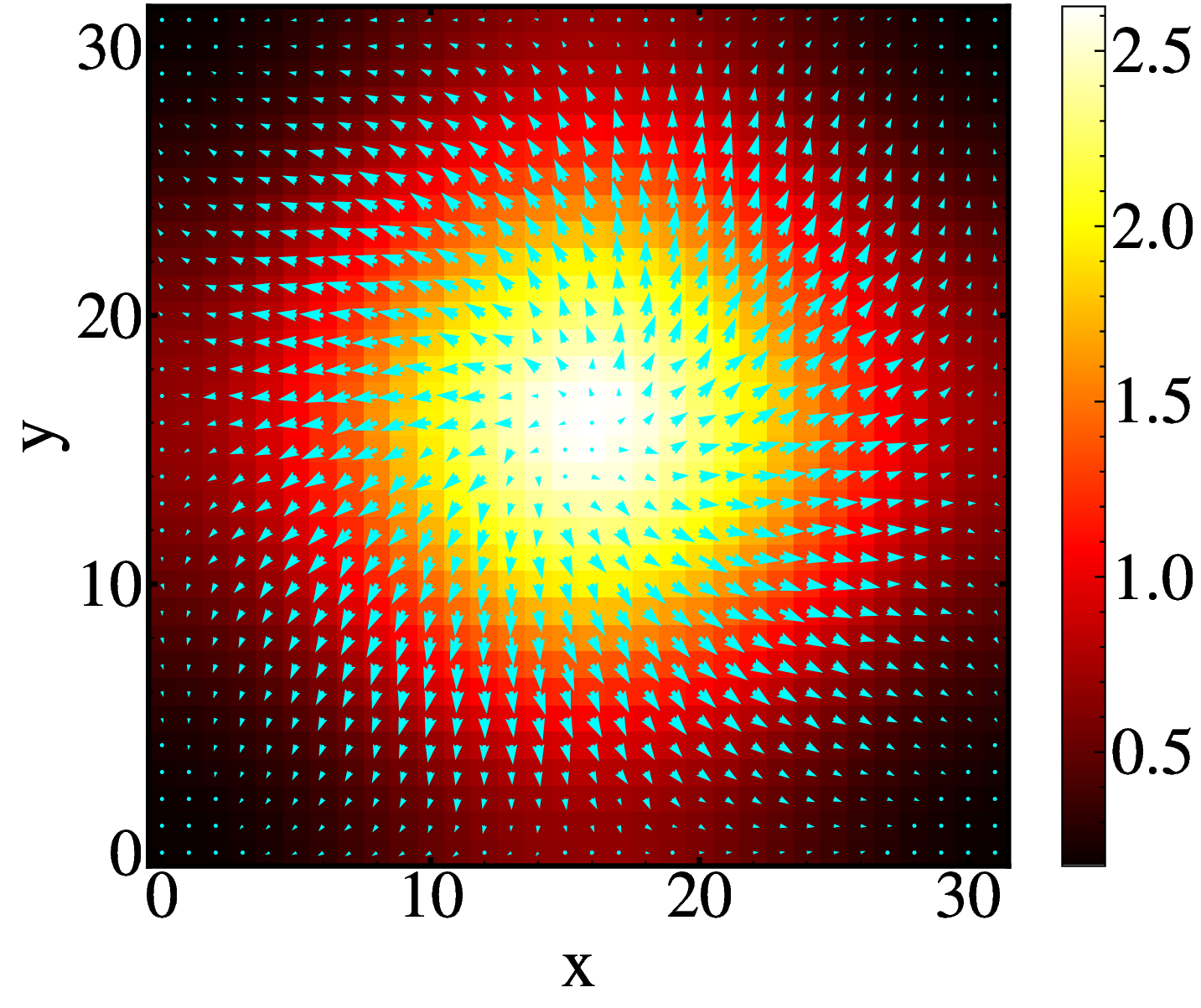}
\put(4,87){\large\bfseries (c)}
\end{overpic}
\end{minipage}

\caption{\textit{Relaxation of a localized density profile on an infinite domain.}
Density (color shade) and current (arrows) profiles at different times:
(a) $T=25$, (b) $T=50$, and (c) $T=100$ for isotropic mass transfer.
The initial condition is a delta-density profile with the entire mass concentrated at the origin; we take global density $\bar \rho=1$, mass chipping (retention) parameters $\lambda_1=0.0$, and $\lambda_2=0.0$. One can see the circulating (transient) currents around the origin. }

\label{fig:relaxation_profiles_isotropy_anisotropy}
\end{figure*}

On the square lattice, we define the density variable $\rho(x,y,t) \equiv \langle m(x,y,t) \rangle$ as the average of local mass; then, the continuity equation can immediately be written in a discrete form by identifying the microscopic currents $\mathbf{j}(x,y,t) = \{ j_x(x,y,t), j_y(x,y) \}$. We first consider the simplest case of hopping dynamics in the bulk, corresponding to the equal hopping probabilities, $p_1 = p_2 = p_3 = p_4 = 1/4$, along all possible directions; as discussed later in Sec. \eqref{edge-curr}, for open boundaries, these rates can however be different depending on the specific boundary conditions imposed at the boundary sites. Indeed, from the infinitesimal-time evolution equation of mass as given in Eq. \eqref{eq:mass_update}, we can readily identify the (instantaneous) diffusive currents in the bulk. The diffusive current along  $x$-direction is given by
\begin{align}
{\rm j}^{({\rm d})}_x &(x,y,t)
= \frac{1}{4} \mu_1 (\tilde{\lambda}_x+ \tilde{\lambda}_{yx})
\left[m(x,y,t)-m(x+1,y,t)\right]  \nonumber \\
&\quad +\frac{1}{4}\mu_1^2 \tilde{\lambda}_y \tilde{\lambda}_{yx}
\Big[m(x,y+1,t)-m(x+1,y-1,t) \Big] 
\end{align}
and, along $y$-direction,
\begin{align}
{\rm j}^{({\rm d})}_y &(x,y,t)
= \frac{1}{4} \mu_1 (\tilde{\lambda}_y+ \tilde{\lambda}_{xy})
\left[m(x,y,t)-m(x,y+1 ,t)\right]  \nonumber \\
&\quad +\frac{1}{4}\mu_1^2 \tilde{\lambda}_x \tilde{\lambda}_{xy}
\Big[m(x-1,y,t)-m(x+1,y+1,t) \Big] .
\end{align}
To obtain a hydrodynamic description of current and denstiy, we perform a gradient expansion of the lattice density field by suitably rescaling  space and time (diffusive scaling). That is, by expanding densities in leading order of gradients, we obtain
\begin{align}
\rho_{x\pm1,y} &= \rho \pm \partial_x \rho + \frac{1}{2}\partial_x^2 \rho + \cdots, \\
\rho_{x,y\pm1} &= \rho \pm \partial_y \rho + \frac{1}{2}\partial_y^2 \rho + \cdots,
\end{align}
and, then by retaining only leading-order terms, the lattice currents reduce to
\begin{equation}
\langle {\rm j}_{\alpha} \rangle = - [\mathbb{D}]_{\alpha \beta}\,\partial_{\beta} \rho.
\end{equation}
In the above equations, we denote $\alpha, \beta \in \{x,y\}$ the cartesian components of a vector, and we define the diffusion tensor as
\begin{equation}
\mathbb{D}
= \frac{1}{4}\mu_1
\begin{pmatrix}
\tilde{\lambda}_x+ \tilde{\lambda}_{yx} + \mu_1\tilde{\lambda}_y \tilde{\lambda}_{yx} & -2 \mu_1 \tilde{\lambda}_y \tilde{\lambda}_{yx} \\
2\mu_1 \tilde{\lambda}_x \tilde{\lambda}_{xy} & \tilde{\lambda}_y+ \tilde{\lambda}_{xy} + \mu_1\tilde{\lambda}_x \tilde{\lambda}_{xy}
\end{pmatrix}.
\end{equation}
Therefore, the time-evolution equation for density field can be written as 
\begin{align}
\partial_t \rho
&=
\sum_{\alpha, \beta} D_{\alpha\beta}\,
\partial_\alpha\partial_\beta \rho
\\
&= D_{xx}\,\partial_x^2 \rho + D_{yy}\,\partial_y^2 \rho + \left(D_{yx} + D_{xy}\right)\,\partial_x \partial_y \rho, 
\label{den-relax}
\end{align}
where the off-diagonal (odd) components of the diffusion tensor enter only through their symmetric combination. In Fig.~\ref{fig:relaxation_profiles_isotropy_anisotropy}, the density and current profiles are plotted at three different times following an initially localized density profile, for $\lambda_1=0$ and $\lambda_2=0$, on an infinite domain. In all three panels (a)--(c), the current (arrow) exhibits a definite handedness, which arises from the chiral mass-transfer rules in the model. Note that the off-diagonal components of the diffusion tensor can, in general, influence the relaxation dynamics. In the special case $D_{xy}=-D_{yx}$, however, the contribution of the odd components to the third term in the density-evolution equation vanishes, and the odd dynamics does not contribute to diffusive relaxation. By contrast, when $|D_{xy}|\neq|D_{yx}|$, the mixed-gradient term [the third term in Eq. \eqref{den-relax}] contributes to the relaxation dynamics.


Interestingly, even when currents are on average absent in the bulk, the odd (off-diagonal) components of the diffusion tensor manifest themselves through an edge current, a hallmark of mirror (parity) symmetry violation. That is, provided that periodic boundary condition is imposed only along the $x$-direction and the transverse direction have reflecting boundaries (``walls''), the system supports edge currents tangential to the wall, which is considered next.

\section{Edge currents}
\label{edge-curr}

In this section, we consider a system with open boundary conditions, with periodicity retained only along the $x$ direction. Along the transverse ($y$) direction, we impose ``impenetrable walls'', preventing mass from escaping and causing it to be reflected back into the bulk. The resulting imbalance of chiral mass transfer generates a steady edge current flowing tangentially to the walls. In this geometry, the boundaries in the transverse direction constitute physical edges of the system, and the edge current can be calculated exactly.

We define the edge current as the net particle current flowing along $x$-direction, measured locally at the boundary sites. Specifically, it is obtained by summing the microscopic hopping currents along the $x$ direction over all sites belonging to a given edge. This quantity characterizes transport localized near the boundary and distinguishes the edge current from the bulk current. We consider two representative choices of mass-transfer rates at the open boundaries, which essentially make the currents perpendicular to the ``wall'' (edge) vanish, and, as we see below, they result in edge currents flowing along the walls (i.e., along $x$ direction).
\\
\\ 
$(i)$ \textbf{Case 1:} Let us first consider a case where we assign the following hopping probabilities at the edges (the sites at the open boundaries); see Fig. \ref{fig:edge current}.(a) and (c) for schematic representation. At the boundary sites with $y=(L-1)$, we set mass transfer probabilities along $-x$ and $-y$ directions as $p_2 = p_4 = 1/4$, respectively; with probability $1-p_2-p_4$, nothing happens (i.e., mass transfers along $+x$ and $+y$ directions are omitted). Likewise, at the boundary sites with $y=0$, we set mass transfer probabilities along $+x$ and $+y$ directions as $p_1 = p_3 = 1/4$, respectively; with probability $1-p_1-p_3$, nothing happens (i.e., mass transfers along $-x$ and $-y$ directions are omitted). 
\\
\\ 
$(ii)$ \textbf{Case 2:} In this case, we assign the following hopping probabilities at the edges; see Fig. \ref{fig:edge current}.(d) and (f) for schematic representation. At the boundary sites $y=(L-1)$, we set mass transfer probabilities along $-x$ and $-y$ directions as $p_2 = p_4 = 1/2$, and mass transfers along $+x$ and $+y$ directions are omitted. At the boundary sites $y=0$, we set $p_1 = p_3 = 1/2$, and mass transfers along $-x$ and $-y$ directions are omitted.  
\\
\\
Note that the mass transfer rates in the bulk are still isotropic: We set $\tilde \lambda_x = \tilde \lambda_y = \tilde \lambda_1$, $\tilde \lambda_{xy}=\tilde\lambda_{yx}=\tilde{\lambda}_2$ and $p_1=p_2=p_3=p_4={1}/{4}$.

In the first case, although the steady-state density profile remains uniform throughout, the current profile is nonuniform, with a nonzero edge current emerging at the boundary sites. The average edge current can be readily calculated in terms of the microscopic parameters $\mu_1$, $\tilde{\lambda}_1$, and $\tilde{\lambda}_2$ as given below:
\begin{align}
    &\langle j_x^{\rm (d)} \rangle(x,y=0) = \frac{1}{4} \mu_1 \Big[ \tilde{\lambda}_1 + \tilde{\lambda}_2 (1+ \mu_1 \tilde{\lambda}_1) \Big] \bar{\rho}\\
    &\langle j_x^{\rm (d)} \rangle (x,y=L-1) = -\frac{1}{4} \mu_1 \Big[ \tilde{\lambda}_1 + \tilde{\lambda}_2 (1+ \mu_1 \tilde{\lambda}_1) \Big] \bar{\rho}.
\end{align}




\begin{figure*}[ht!]
\centering
\begin{minipage}{0.485\linewidth}
\centering
\begin{overpic}[width=\linewidth]{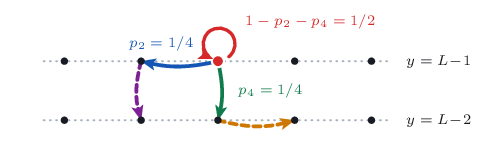}
\put(1,25){\large\bfseries (a)}
\end{overpic}\\[3pt]
\begin{overpic}[width=\linewidth]{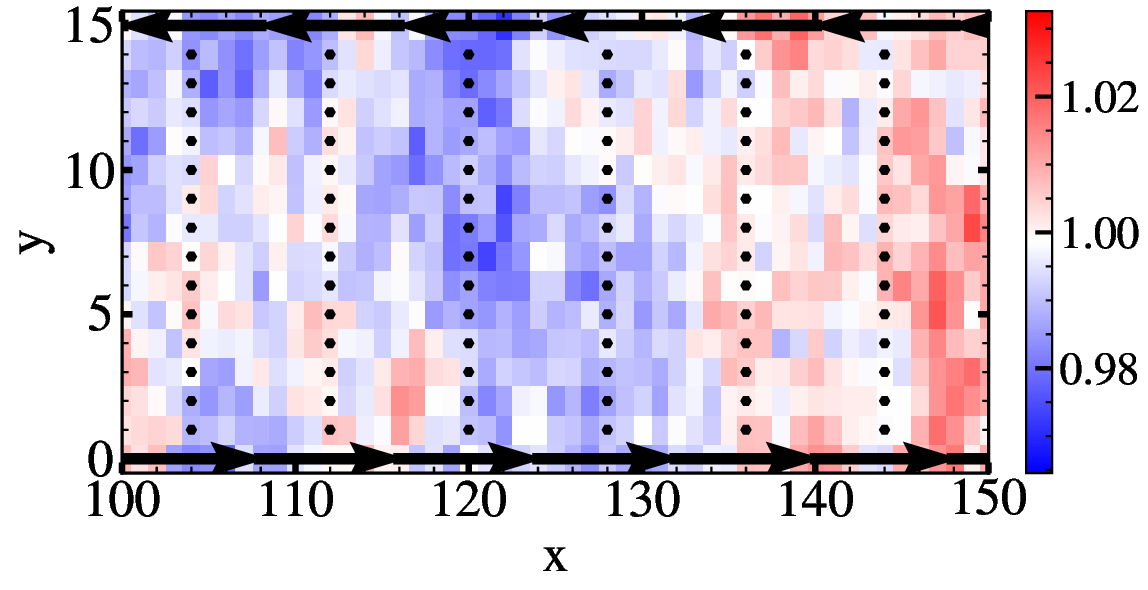}
\put(4,54){\large\bfseries (b)}
\end{overpic}\\[3pt]
\begin{overpic}[width=\linewidth]{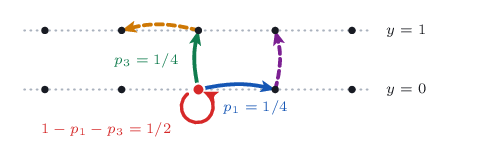}
\put(1,25){\large\bfseries (c)}
\end{overpic}
\end{minipage}
\hfill
\begin{minipage}{0.485\linewidth}
\centering
\begin{overpic}[width=\linewidth]{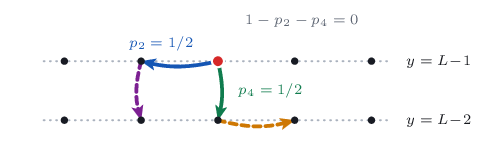}
\put(1,25){\large\bfseries (d)}
\end{overpic}\\[3pt]
\begin{overpic}[width=\linewidth]{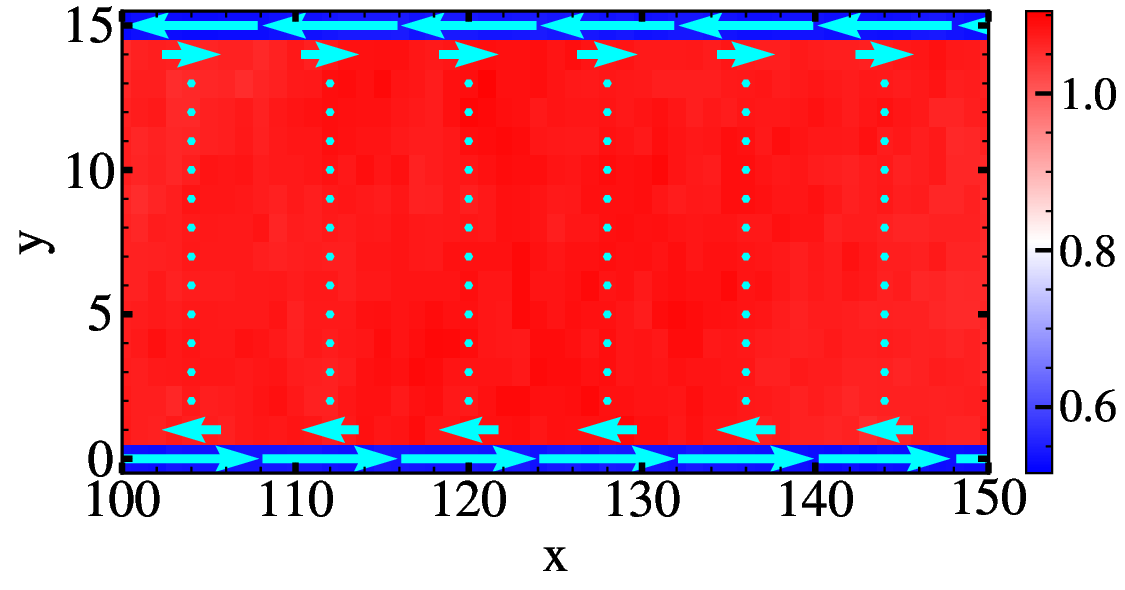}
\put(4,54){\large\bfseries (e)}
\end{overpic}\\[3pt]
\begin{overpic}[width=\linewidth]{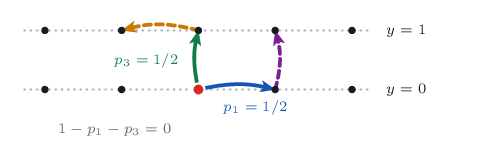}
\put(1,25){\large\bfseries (f)}
\end{overpic}
\end{minipage}
\caption{\textit{Steady-state edge currents under open-boundary conditions.}
Panels (b) and (e) show heat maps of the steady-state density profile, with arrows indicating the current profiles. The schematic diagram above and below each of the heat maps give the corresponding mass-transfer rules at the two reflecting ``walls'' (edges), placed adjacent to $y=(L-1)$ [panels (a) and (d)] and $y=0$ [panels (c) and (f)]. Of the four chiral ``L''-shaped moves, only the two moves whose path stays inside the slab survive at the edge (wall): $p_2,p_4$ at $y=(L-1)$, so that $p_1=p_3=0$, and $p_1,p_3$ at $y=0$, so that $p_2=p_4=0$. In panels (a) and (c), the surviving moves keep their bulk value $1/4$, and with the remaining probability $1/2$ no mass is transferred at all (red self-loop); in panels (d) and (f), they are raised to $1/2$, so every event transfers mass. Although the density remains nearly homogeneous in the bulk, pronounced currents persist near the edges. Indeed, their qualitative structure indicates that the observed edge transport is a robust manifestation of the underlying chiral dynamics. We take the following simulation parameters: system size
$L_x\times L_y=512\times16$, global density $\bar \rho=1$, $\lambda_1=0.4$, $\lambda_2=0.6$, and $\mu_1=1/2$.}
\label{fig:edge current}
\end{figure*}

In the second case, the density profile is modified near the boundaries and rapidly approaches its bulk value away from them. Denoting the bulk density by $\bar{\rho}$, the boundary density $\rho_0$ is given by
\begin{align}
    \rho_{0} = \frac{\bar{\rho}}{2} .
\end{align}
Consequently, the average edge currents at the two open boundaries are given by
\begin{align}
    &\langle j_x^{\rm (d)} \rangle (x, y=0) = \frac{1}{4} \mu_1 \Big[ \tilde{\lambda}_1 + \frac{1}{2}\tilde{\lambda}_2 + \mu_1 \tilde{\lambda}_1 \tilde{\lambda}_2 \Big] \bar{\rho}\\
    &\langle j_x^{\rm (d)} \rangle (x, y=L-1) = -\frac{1}{4} \mu_1 \Big[ \tilde{\lambda}_1 + \frac{1}{2}\tilde{\lambda}_2 + \mu_1 \tilde{\lambda}_1 \tilde{\lambda}_2 \Big] \bar{\rho}.
\end{align}
The currents in the adjacent (boundary) layers are correspondingly given by
\begin{align}
    &\langle j_x^{\rm (d)} \rangle  (x,y = 1)= - \frac{1}{4} \mu_1 \tilde{\lambda}_2 \bar{\rho}, \\
    &\langle j_x^{\rm (d)} \rangle  (x,y= L-2) = \frac{1}{4} \mu_1 \tilde{\lambda}_2 \bar{\rho}.
\end{align}
In Fig. \eqref{fig:edge current}, heat maps of the steady-state current and density profiles on the $x$-$y$ plane are shown for the following two cases: {\it Case 1.} with the boundary condition at $y=L-1$ in panel (a) and at $y=0$ in panel (c), and {\it Case 2.} with the boundary conditions  at $y=L-1$ in panel (d) and  at $y=0$ in panel (f). These cases correspond to the current profiles presented in panels (b) and (e), respectively. Notably, although the current and density profiles in both cases remain localized near the boundaries, their detailed spatial structures depend on the specific mass-transfer rates imposed at the boundary sites.

\section{Current fluctuations}
\label{Curr-fluc}

In this section, we characterize the steady-state current fluctuations in the bulk. To this end, we consider periodic boundary conditions in both the $x$ and $y$ directions. In lattice models with conserved mass, transport occurs through stochastic mass transfers across bonds connecting neighboring sites. A natural quantity for characterizing such transport is the time-integrated bond current, defined as the net flux of mass across a given bond over a finite time interval. Specifically, if an amount of mass is transferred across a bond along positive direction, we add it to the cumulative current and, if it is transferred along negative direction, we subtract the transferred mass from the cumulative current. Due to the stochastic nature of the dynamics, the instantaneous mass and current fluctuate around its mean. Here we focus on their steady-state fluctuation properties.

We characterize the fluctuation and relaxation properties of these chiral mass-transport processes in terms of the two-point spatial density correlation function (equivalently, the structure factor), the bulk-diffusion coefficients and the Onsager transport coefficients (or, equivalently, the mobility).
Let $Q_x(x,y,t)$ and $Q_y(x,y,t)$ denote the net mass flux in a time interval $[0,t]$ across bonds $[(x,y),(x+1, y)]$ and $[(x, y),(x,y+1)]$, respectively. That is, we can write
\begin{equation}
 Q_{\alpha}(x,y,t) = \int_{0}^{t} dt' \, j_{\alpha}(x,y, t'),
\end{equation}
where $j_{\alpha}(x,y, t')$ is the instantaneous current along $\alpha$th principal axis at time $t'$. In the present study, we focus on the steady-state fluctuation properties of the current variables.
On a periodic lattice, we have the first moments of current vanishing in the steady state, i.e., $\langle Q_{\alpha} \rangle =0$ and $\langle j_{\alpha} \rangle=0$.

The cumulative, or time-integrated, bond current is a fundamental observable in nonequilibrium systems, as it quantifies the transport statistics of mass through the system over a finite time interval. In contrast to the instantaneous current, whose fluctuations are strongly influenced by microscopic fluctuations (noise), the fluctuations of $Q_{\alpha}$ exhibit well-defined scaling behavior in the long-time limit. Indeed, the variance, and higher-order cumulants, of $Q_{\alpha}$ encode important transport characteristics of the system.
Another quantity of interest is the cumulative (space-time-integrated) current across the whole system, which characterizes macroscopic transport properties of the system.
In particular, the long-time growth rate of the variance of the space-time-integrated current, as discussed later, determines the mobility that enters fluctuating hydrodynamics and macroscopic fluctuation theory \cite{Bertini2001Jul, Bertini2015Jun}. The statistics of the space-time-integrated  current thus provide a direct probe of nonequilibrium transport, which we characterize below at the level of two-point correlation functions.

In particular, the equal-time correlations of the time integrated bond current,
\begin{align}
    C_{r,s}^{Q_{\alpha}Q_{\beta}}(t,t) = \langle Q_{\alpha}(r,s,t) Q_{\beta}(0,0,t)\rangle_c,
\end{align}
satisfy time-evolution equations, whose source terms are encoded in a density-dependent transport coefficient, a tensor $\Gamma^{\alpha\beta}_{r,s}$, that depends on the microscopic details and is quantified below. To this end, it is physically quite instructive to make a decomposition of bond current, 
$$
j_{\alpha}(x,y,t) = j_{\alpha}^{(\rm d)}(x,y,t) + j_{\alpha}^{(\rm fl)}(x,y,t),
$$ 
into a slow (hydrodynamic) and fast (``noise'') variables, respectively. Then, using the infinitesimal-time update rules as in Eqs. \eqref{curr-update1} and \eqref{curr-update2}, we can write the time-evolution of the current correlation function in terms of the above-mentioned current variables as 
\begin{align}
\label{current-cor}
    \partial_t C_{r,s}^{Q_{\alpha}Q_{\beta}}(t,t) = C_{r,s}^{j^{(d)}_{\alpha}Q_{\beta}}(t,t) + C_{r,s}^{Q_{\alpha}j^{(d)}_{\beta}}(t,t) + \Gamma_{r,s}^{\alpha,\beta}.
\end{align}
Indeed, the quantity $\Gamma^{\alpha\beta}_{r,s}$ can be shown to be directly related to the fluctuating (``noise'') part of the instantaneous current  through the following correlations:
\begin{equation}
\langle j_{\alpha}^{(\rm fl)}(r,s,t)
j_{\beta}^{(\rm fl)}(0,0,t')
\rangle
=
\Gamma^{\alpha\beta}_{r,s}\,
\delta(t-t').
\label{eq:current_noise}
\end{equation}
The density-dependent quantity $\Gamma^{\alpha\beta}_{r,s}$—the “microscopic” mobility tensor—characterizes the strength of the fluctuating (``noise'') current and, notably, is {\it not} delta-correlated in space. While the temporal delta function indicates that the noise current is white in time, the spatial dependence of $\Gamma^{\alpha \beta}_{r,s}$ however encodes short-ranged correlations arising from the underlying mass-transport dynamics, which involves coordinated, simultaneous hopping across multiple spatial locations (bonds) on the lattice. Moreover, $\Gamma^{\alpha\beta}_{r,s}$ enters as the source term in the evolution equations for the integrated-current correlations and therefore governs their long-time growth. 
As we show below, the spatial structure of these noise-current correlations—in particular, the long-wavelength structure of the correlations between the orthogonal current components (off-diagonal elements of the mobility tensor)—induces a nonanalytic contribution to the static structure factor. This nonanalyticity, in turn, gives rise to algebraically decaying two-point density correlations for the {\it chiral}, albeit {\it isotropic}, mass-conserving dynamics studied in the present work.

For the chiral model considered here, components of the mobility tensor $\Gamma^{\alpha \beta}_{r,s}$ can be explicitly written in terms of the microscopic parameters as given below:
\begin{subequations}
\begin{align}
\Gamma^{xx}_{r,s}
&=
\Bigg[
\frac{\mu_2}{2}
\left(
\tilde{\lambda}_x^2
+\tilde{\lambda}_{yx}^2
+\mu_2\tilde{\lambda}_y^2\tilde{\lambda}_{yx}^2
\right)M_0
\nonumber\\
&\qquad 
+\mu_1\mu_2
\tilde{\lambda}_y
\tilde{\lambda}_{yx}^2
M_{1y}
\Bigg]
\delta(r)\delta(s),
\\[2mm]
\Gamma^{yy}_{r,s}
&=
\Bigg[
\frac{\mu_2}{2}
\left(
\tilde{\lambda}_y^2
+\tilde{\lambda}_{xy}^2
+\mu_2\tilde{\lambda}_x^2\tilde{\lambda}_{xy}^2
\right)M_0
\nonumber\\
&\qquad
+\mu_1\mu_2
\tilde{\lambda}_x
\tilde{\lambda}_{xy}^2
M_{1x}
\Bigg]
\delta(r)\delta(s),
\\[2mm]
\Gamma^{xy}_{r,s}
&=
\frac{\mu_1}{4}
\tilde{\lambda}_x
\tilde{\lambda}_{xy}
\left(
\mu_1M_{1x}
+\mu_2\tilde{\lambda}_xM_0
\right)
\nonumber\\
&\qquad\times
\Big[
\delta(r+1)\delta(s)
+
\delta(r)\delta(s-1)
\Big]
\nonumber\\
&\quad
-\frac{\mu_1}{4}
\tilde{\lambda}_y
\tilde{\lambda}_{yx}
\left(
\mu_1M_{1y}
+\mu_2\tilde{\lambda}_yM_0
\right)
\nonumber\\
&\qquad\times
\Big[
\delta(r+1)\delta(s-1)
+
\delta(r)\delta(s)
\Big],
\\[2mm]
\Gamma^{yx}_{r,s}
&=
\frac{\mu_1}{4}
\tilde{\lambda}_x
\tilde{\lambda}_{xy}
\left(
\mu_1M_{1x}
+\mu_2\tilde{\lambda}_xM_0
\right)
\nonumber\\
&\qquad\times
\Big[
\delta(r-1)\delta(s)
+
\delta(r)\delta(s+1)
\Big]
\nonumber\\
&\quad
-\frac{\mu_1}{4}
\tilde{\lambda}_y
\tilde{\lambda}_{yx}
\left(
\mu_1M_{1y}
+\mu_2\tilde{\lambda}_yM_0
\right)
\nonumber\\
&\qquad\times
\Big[
\delta(r-1)\delta(s+1)
+
\delta(r)\delta(s)
\Big].
\end{align}
\label{eq:Gamma matrix coefficients}
\end{subequations}
Here, the undetermined density-dependent constants,
\begin{align}
M_0 &= \langle m^2(x,y)\rangle,\notag\\
M_{1x} &= \langle m(x,y)m(x+1,y)\rangle,\notag\\
M_{1y} &= \langle m(x,y)m(x,y+1)\rangle,\notag
\end{align}
represent the steady-state second moment and nearest-neighbor correlations of local masses: $M_0$ denote the onsite mass fluctuation and $M_{1x}$ and $M_{1y}$ denote the nearest-neighbor correlations along the $x$ and $y$ directions, respectively ($M_{1x} = M_{1y} \equiv M_1$ for isotropic hopping rates). These quantities can be determined self-consistently from the steady-state mass correlation functions; see Eqs. \eqref{cond1} and \eqref{cond2} for detailed discussion.

For equilibrium systems, it is well known from the linear-response theory that, on the large spatio-temporal scales, the variance of space-time-integrated current can be characterized by the hydrodynamic mobility tensor through the familiar Green-Kubo fluctuation-response relation. For nonequilibrium systems that violate detailed balance, however, the validity of such Green-Kubo-like relations are far from obvious. Next, we investigate this question for the chiral mass-transport processes considered here. To this end, we introduce the space-time-integrated current $\bar{Q}_{\alpha }(A,T)$, defined as
\begin{align}
    \bar{Q}_{\alpha}(A,T) = \sum_{x=0}^{L_x -1} \sum_{y=0}^{L_y -1} Q_{\alpha}(x,y,T)\\=  \sum_{x=0}^{L_x -1} \sum_{y=0}^{L_y -1} \int_{0}^{T}d \tau \, j_{\alpha}(x,y,\tau)\\
    = \sum_{x=0}^{L_x -1} \sum_{y=0}^{L_y -1} \int_{0}^{T}d \tau \, j_{\alpha}^{(\rm fl)}(x,y,\tau),
\end{align}
where $\alpha \in \{ x,y\}$ and $A=L_x L_y$ is the total area (in general volume in higher dimensions). In the last step of the above equation, we have used that $\sum_{x,y}j_{\alpha}^{(\rm d)}=0$, which is the consequence of the models being of ``gradient-type'' \cite{Bertini2015Jun}. Then we define the ``macroscopic'' (hydrodynamic) mobility in terms of the scaled variance of space-time-integrated current as
\begin{align}\label{eq:chi_ab}
    \chi^{\alpha \beta} = \lim_{T\to \infty A \to \infty} \frac{1}{2AT} \Big \langle \bar{Q}_{\alpha}(A,T) \bar{Q}_{\beta}(A,T)  \Big\rangle.
\end{align}
Now, by substituting the expression of $\bar{Q}_{\alpha}(A,T)$ in the above equation and using Eq.~(\ref{eq:current_noise}), we can write the mobility tensor as the correlation between the respective currents in directions $\alpha$ and $\beta$,
\begin{align}
     \nonumber \chi^{\alpha \beta} = &\lim_{A,T \to \infty} \frac{1}{2AT} \sum_{x,y}\sum_{x',y'} \\
     &\times \int_{0}^{T} dt \int_{0}^{T} dt' \Big \langle j_{\alpha}^{(\rm fl)}(x,y,t) j_{\beta}^{(\rm fl)}(x',y',t')  \Big \rangle \\
    =\lim_{A,T \to \infty} &\frac{1}{2AT} \sum_{x,y} \sum_{x',y'} \Gamma^{\alpha \beta}_{x-x',y-y'} \int_{0}^{T} dt \int_{0}^{T} dt' \delta(t-t') \\
    = \lim_{A,T \to \infty} &\frac{1}{2AT} \Big( A \sum_{r,s}\Gamma_{r,s}^{\alpha \beta} \Big) T  = \frac{1}{2}\sum_{r,s}\Gamma_{r,s}^{\alpha\beta}.
    \label{eq:hydrodynamic_mobility}
\end{align}
That is, we obtain a nonequilibrium version of the familiar equilibrium Green-Kubo relation,
\begin{align}\label{eq:chi_gamma}
    \chi^{\alpha \beta} = \frac{1}{2}\sum_{r,s} \Gamma_{r,s}^{\alpha\beta},
\end{align}
which expresses the ``macroscopic'' (hydrodynamic) mobility tensor in terms of the strength of noise-current correlation, $\Gamma_{r,s}^{\alpha\beta}$, which has been calculated exactly for the chiral systems considered in the present work.
Thus, the hydrodynamic mobility tensor can be obtained directly from the variance of the spatial integral of current in the system, connecting the microscopic fluctuations and the emergent macroscopic transport coefficients.
Now, using Eq.~(\ref{eq:hydrodynamic_mobility}), we can explicitly write the components of the hydrodynamic mobility tensor as
\begin{align}
\chi^{xx}
&=
\frac{1}{4}
\Big[
\mu_2
\left(
\tilde{\lambda}_x^2
+\tilde{\lambda}_{yx}^2
+\mu_2 \tilde{\lambda}_y^2 \tilde{\lambda}_{yx}^2
\right)M_0
\nonumber\\
&\qquad
+
2\mu_1\mu_2
\tilde{\lambda}_y
\tilde{\lambda}_{yx}^2
M_{1y}
\Big],
\\
\chi^{yy}
&=
\frac{1}{4}
\Big[
\mu_2
\left(
\tilde{\lambda}_y^2
+\tilde{\lambda}_{xy}^2
+\mu_2 \tilde{\lambda}_x^2 \tilde{\lambda}_{xy}^2
\right)M_0
\nonumber\\
&\qquad
+
2\mu_1\mu_2
\tilde{\lambda}_x
\tilde{\lambda}_{xy}^2
M_{1x}
\Big],
\\
\chi^{xy}
=
\chi^{yx}
&=
\frac{\mu_1}{4}
\Big[
\tilde{\lambda}_x
\tilde{\lambda}_{xy}
\left(
\mu_1 M_{1x}
+
\mu_2 \tilde{\lambda}_x M_0
\right)
\nonumber\\
&\qquad
-
\tilde{\lambda}_y
\tilde{\lambda}_{yx}
\left(
\mu_1 M_{1y}
+
\mu_2 \tilde{\lambda}_y M_0
\right)
\Big].
\label{eq:mobility for general model}
\end{align}
Next, we consider a special case--the chiral models with isotropic mass-transfer rates.

\section{Isotropic Mass Transfer}
\label{iso-mcm}

In the rest of the paper, we confine ourselves to chiral mass-transport processes on a periodic domain (i.e., no edge in the systems) with {\it isotropic} hopping dynamics in the bulk.
Isotropic systems with broken parity symmetry have recently attracted considerable interest in a variety of contexts, ranging from chiral active matter and Hall transport to odd hydrodynamics \cite{Lowen2016Nov, Fruchart2023Mar}. They provide a minimal setting in which conventional diffusive relaxation coexists with parity-breaking transport. Accordingly, we consider the symmetric choice of the microscopic (bulk) parameters: Mass-chipping fractions are set to
\begin{equation}\label{eq:lambda_param}
\tilde{\lambda}_x=\tilde{\lambda}_y=\tilde{\lambda}_1,
\qquad
\tilde{\lambda}_{xy}=\tilde{\lambda}_{yx}=\tilde{\lambda}_2,
\end{equation}
and directional hopping probabilities are set to
\begin{equation}\label{eq:probs_iso}
p_1=p_2=p_3=p_4=\frac{1}{4},
\end{equation}
in the bulk.

\subsection{Diffusion tensor}

This particular choice restores rotational invariance (isotropy), while preserving the chiral character of the microscopic dynamics. The resulting bulk-diffusion tensor takes the following form:
\begin{equation}
\mathbb{D} =
\begin{pmatrix}
D & -D^{(\rm odd)}\\
D^{(\rm odd)} & D
\end{pmatrix},
\end{equation}
where we can explicitly obtain the individual elements of the tensor as 
\begin{align}
D_{xx}=D_{yy}
&\equiv D
=
\frac{1}{4}
\Big[
\mu_1(\tilde{\lambda}_1+\tilde{\lambda}_2)
+
\mu_1^2\tilde{\lambda}_1\tilde{\lambda}_2
\Big],
\\
D_{xy}=-D_{yx}
&\equiv -D^{(\rm odd)}
=
\frac{1}{2}
\mu_1^2
\tilde{\lambda}_1
\tilde{\lambda}_2.
\end{align}
From the perspective of linear irreversible thermodynamics, the coefficient $D$ plays the role of the conventional  bulk-diffusion coefficient and governs dissipative relaxation in the system. In contrast, the quantity $D^{(\rm odd)}$ is an antisymmetric transport coefficient generated by the chiral microscopic dynamics. Such parity-odd transport coefficients are analogous to Hall-type response coefficients and give rise to transverse currents without directly contributing to dissipation.
The diffusion tensor can then be decomposed into symmetric and antisymmetric parts,
\begin{equation}
\mathbb{D}
=
\mathbb{D}_S+\mathbb{D}_A
=
D\,\mathbb{I}
+
D^{(\rm odd)}\,\mathbb{E},
\end{equation}
where the identity matrix $\mathbb{I}$ and the two-dimensional Levi-Civita permutation tensor $\mathbb{E}$ are given by
\begin{equation}
\mathbb{I}
=
\begin{pmatrix}
1 & 0\\
0 & 1
\end{pmatrix},
\qquad
\mathbb{E}
=
\begin{pmatrix}
0 & -1\\
1 & 0
\end{pmatrix}.
\end{equation}
The symmetric contribution $\mathbb{D}_S=D\mathbb{I}$ governs the irreversible spreading of mass, whereas the antisymmetric contribution $\mathbb{D}_A=D^{(\rm odd)}\mathbb{E}$ generates transverse transport associated with chirality.
Equivalently, the diffusive current can be written as
\begin{align}
\mathbf{j}
&= -\mathbb{D} \nabla \rho
\equiv
-\mathbb{D}_S\nabla\rho
-\mathbb{D}_A\nabla\rho
\nonumber \\
&= -D\,\nabla\rho -
D^{(\rm odd)}
\,\hat{\mathbf z}\times\nabla\rho .
\end{align}
The second term represents an odd-diffusive current that is everywhere perpendicular to the density gradient. 
This odd current does not contribute directly to dissipation. Instead, it redistributes mass transverse to the density gradient and can generate circulating current around density inhomogeneities (see Fig.~\ref{fig:relaxation_profiles_isotropy_anisotropy}) while preserving the diffusive relaxation rate set by $D$.

\subsection{Mobility tensor}

Now, we turn to the second transport coefficient, namely the (microscopic) mobility tensor. Using (\ref{eq:Gamma matrix coefficients}), the microscopic fluctuating current covariance tensor can be expressed in Fourier space as
\begin{align}
\Gamma^{\alpha\alpha}_{q_x,q_y}(\bar{\rho})
&= \gamma_0(\bar{\rho}),
\\
\Gamma^{\alpha\beta}_{q_x,q_y}(\bar{\rho})
&= \gamma_2(\bar{\rho})
\left(1-e^{-iq_\alpha}\right)
\left(1-e^{iq_\beta}\right),
\qquad (\alpha\neq\beta).
\end{align}
Here we have introduced the following density-dependent coefficients,
\begin{align}
\gamma_0(\bar{\rho})
&= \frac{1}{2}\mu_2
\Big(
\tilde{\lambda}_1^2
+\tilde{\lambda}_2^2
+\mu_2\tilde{\lambda}_1^2\tilde{\lambda}_2^2
\Big)M_0
+\mu_1\mu_2
\tilde{\lambda}_1
\tilde{\lambda}_2^2
M_1,
\label{g0}\\
\gamma_2(\bar{\rho})
&=
-\frac{1}{4}
\mu_1
\tilde{\lambda}_1
\tilde{\lambda}_2
\Big(
\mu_1M_1
+
\mu_2\tilde{\lambda}_1M_0
\Big),
\label{g2}
\end{align}
which characterize the diagonal and off-diagonal components of the mobility tensor, respectively.
The hydrodynamic mobility tensor follows directly from Eq.~(\ref{eq:mobility for general model}) and, due to the isotropy in the problem, it is proportional to an identity matrix,
\begin{equation}
\boldsymbol{\chi}
=
\frac{1}{2}\gamma_0(\bar{\rho})
\begin{pmatrix}
1 & 0\\
0 & 1
\end{pmatrix} = \frac{1}{2}\gamma_0(\bar{\rho}) \mathbb{I}.
\label{eq:isotropic_mobility}
\end{equation}
Thus, although the microscopic current fluctuations possess nontrivial parity-breaking cross correlations through the microscopic mobility tensor $\Gamma^{\alpha\beta}_{\bf q}$ ($\alpha \ne \beta$), these contributions vanish upon spatial integration. The macroscopic mobility therefore remains isotropic and is completely determined by the coefficient $\gamma_0(\bar{\rho})$; however, for more general hopping rates, it can have nonzero off-diagonal components.
In Fig. \ref{fig:mobility}, we verify, in the case of isotropic chiral model, the current-current correlations (both the variance and the cross-correlations) as mentioned in Eqs. \eqref{eq:chi_ab}, \eqref{eq:chi_gamma} and \eqref{eq:isotropic_mobility}. Here, points are obtained from simulations and lines are obtained from theory [Eq. \eqref{eq:chi_gamma}] with isotropic mass-transfer rates as mentioned in Eqs. \eqref{eq:lambda_param} and \eqref{eq:probs_iso}.

\begin{figure}[h!]
    \centering
    \includegraphics[width=0.9\linewidth]{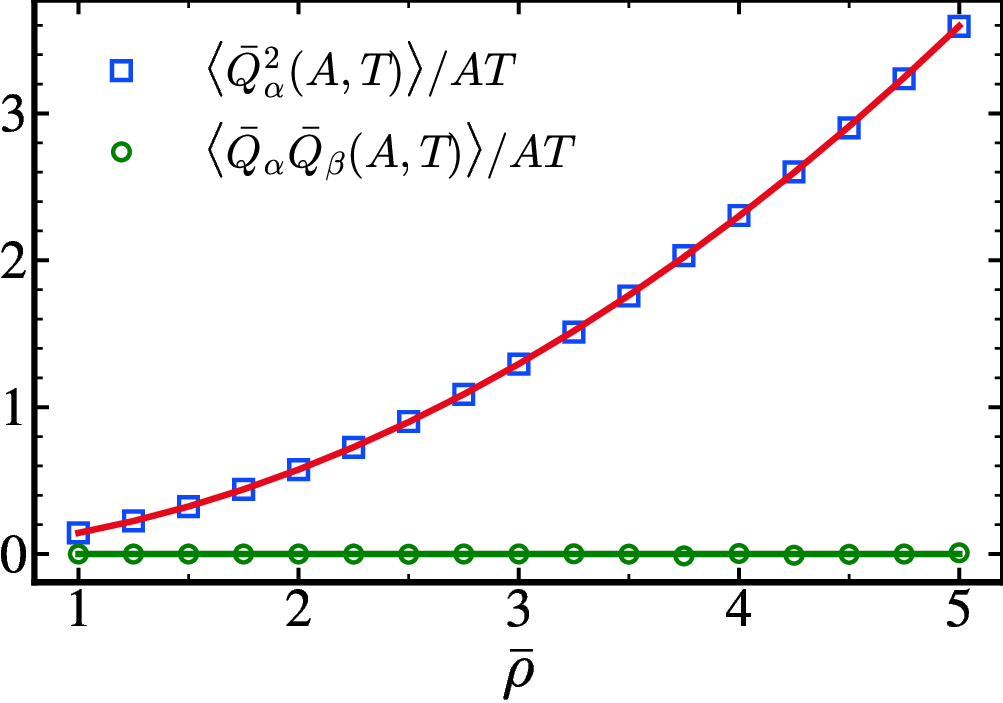}
    \caption{ 
     Scaled steady-state variance of the space-time-integrated (cumulative) current in the full system over the time interval $[0,T]$ are plotted as a function of global density $\rho$ for the isotropic chiral MCM. Open blue squares denote the scaled variance of the cumulative current, while open green circles denote the cross correlation between the two current components. The red line shows the theoretical prediction $2\chi^{\alpha\alpha}$ obtained from Eqs. \eqref{g0} and \eqref{eq}, and the green line shows $2\chi^{\alpha\beta}$ ($\alpha\neq\beta$) from Eq. \eqref{eq}. The simulation parameters are as follows: System size $L_x\times L_y=128\times128$, final time $T=500$, mass chipping (retention) parameters $\lambda_1=0.4$, and $\lambda_2=0.6$. 
     }
    \label{fig:mobility}
\end{figure}

\subsection{Density correlations}\label{subsec:dens_corl}


In this section, we investigate mass fluctuations in chiral mass-transport processes with isotropic mass transfer by characterizing the two-point spatial density correlation function $C^{mm}_{r,s}$ in the steady state. Starting directly from the microscopic dynamical rules, we derive the exact time-evolution equation for the equal-time density–density correlation function and analyze its relaxation toward the steady state. Remarkably, the resulting equation closes exactly at the two-point level, and no Bogoliubov–Born–Green–Kirkwood–Yvon (BBGKY) hierarchy of coupled equations involving higher-order correlation functions arises. This closure is a direct consequence of the linear dependence of the microscopic mass currents on the local mass variables: Upon obtaning the microscopic mass-current correlations, the time evolution of a two-point correlation eventually involves only one- and two-point correlations of local masses, with no higher-order correlators being generated. Thus, the correlation dynamics can be solved exactly without invoking any factorization or other closure approximation.


We proceed by performing a discrete Fourier transformation and obtain an exact expression for the steady-state structure factor in wave-vector space, which allows us to identify the scaling behavior and symmetry properties of the fluctuations across different length scales. Finally, by taking the hydrodynamic limit, corresponding to long wavelengths and long times, we extract the large-scale form of the two-point density correlation function and analyze its asymptotic behavior in the steady state.

By using the infinitesimal-time update rules for masses as provided in Eq. \eqref{eq:mass_update}, the resulting time-evolution equation for the density correlation takes the form 
\begin{figure*}
\centering

\begin{minipage}{0.485\linewidth}
\centering
\begin{overpic}[width=\linewidth]{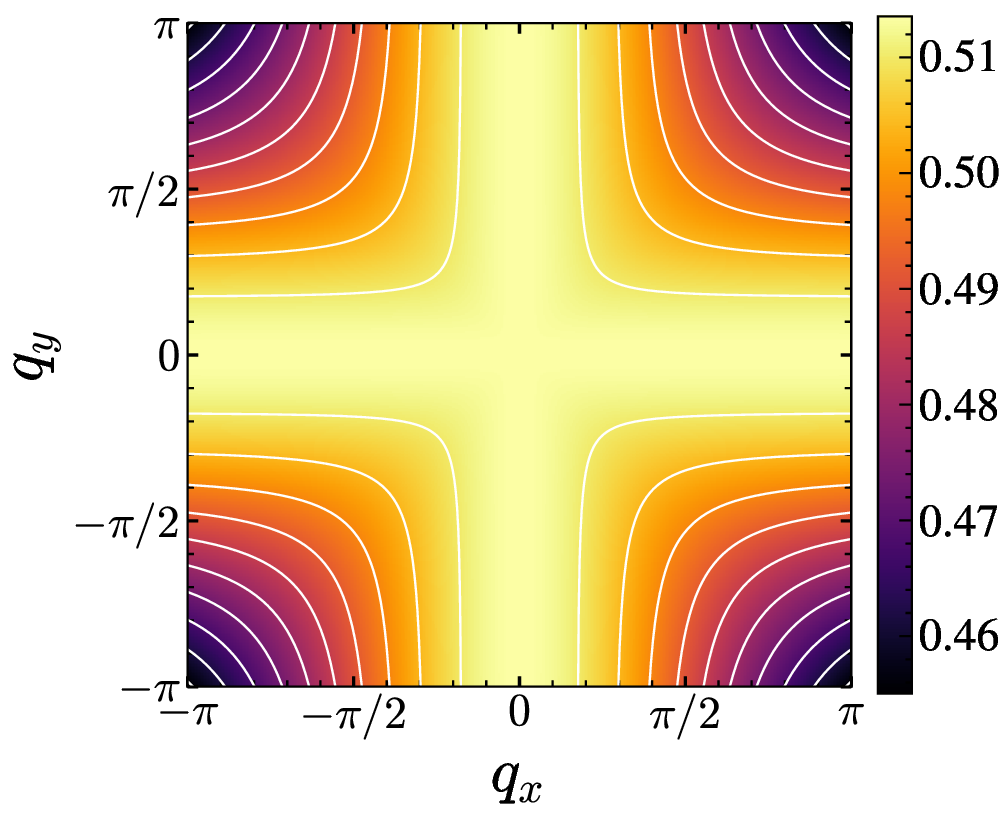}
\put(4,80){\large\bfseries (a)}
\end{overpic}
\end{minipage}
\hfill
\hfill
\begin{minipage}{0.485\linewidth}
\centering
\begin{overpic}[width=\linewidth]{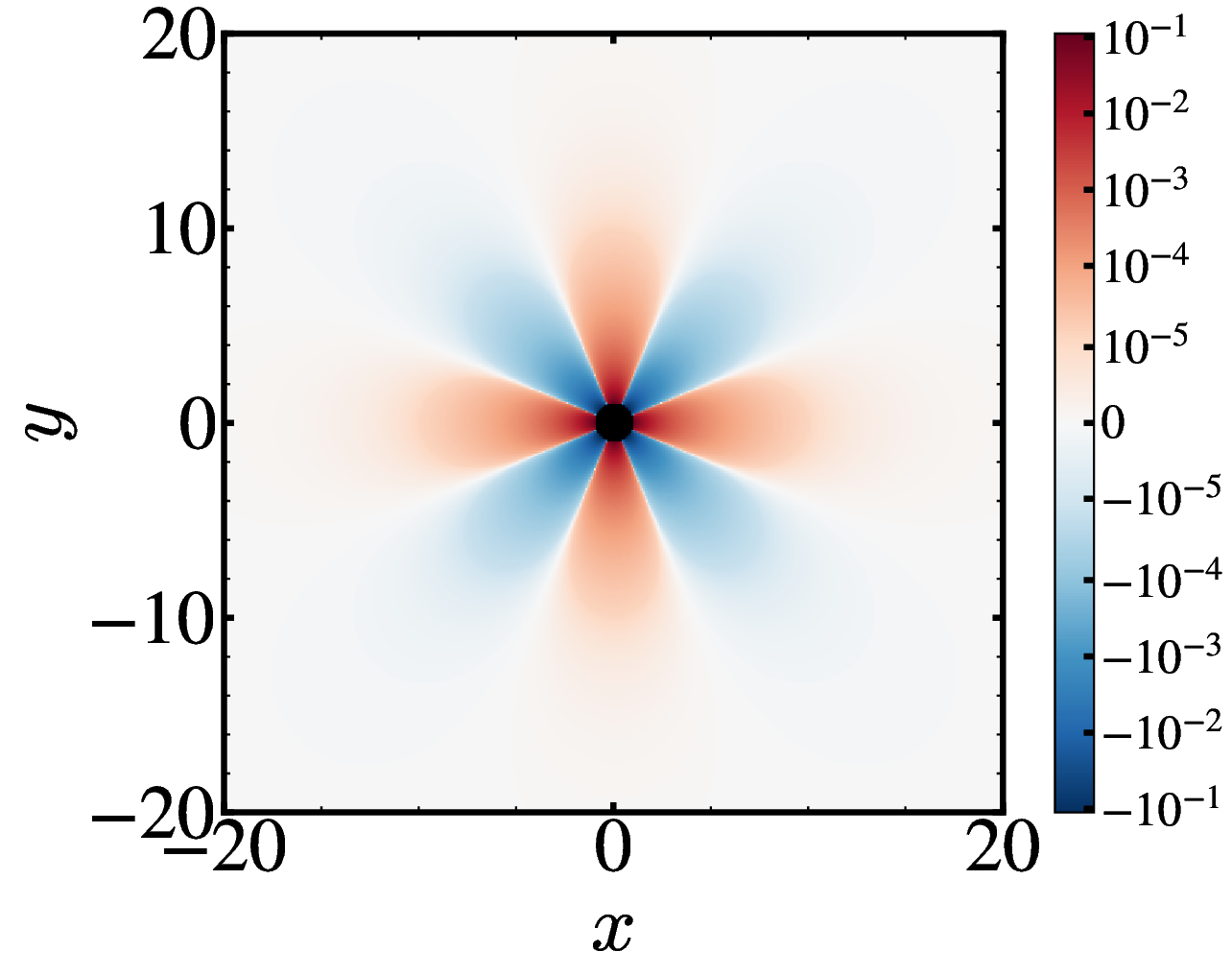}
\put(4,80){\large\bfseries (b)}
\end{overpic}
\end{minipage}
\hfill

\caption{\textit{Steady-state static structure factor and equal-time two-point density correlation function in chiral mass-transport models with isotropic mass-transfer rates.}
Panel (a): Heat map of the structure factor $S(q_x,q_y)$ in wave vector space ${\bf q} =\{q_x,q_y\}$.
Panel (b): Heat map of the corresponding two-point density correlation in real space ${\bf r}=\{x,y\}$.
Here, the simulation parameters are as follows: global density $\bar \rho=1$ and mass chipping (retention) parameters $\lambda_1=0.4$, and $\lambda_2=0.6$.}

\label{fig:structure factors}
\end{figure*}

\begin{align}
\nonumber \partial_t C^{mm}_{r,s} &= 2(D-D^{odd}) \sum_{a=\pm1}\big(C^{mm}_{r+a,s}+C^{mm}_{r,s+a}
-2C^{mm}_{r,s}\big)\\
&+ D^{odd} \sum_{a=\pm1}\sum_{b=\pm1}
\big(
C^{mm}_{r+a,s+b}
-
C^{mm}_{r,s}
\big)
+
B_{r,s}.
\label{den-cor}
\end{align}
where we define a source term
\begin{align}
\nonumber B_{r,s} =& 4(\gamma_0 + 2\gamma_2)\,\delta_{r,0}\delta_{s,0} + 2\gamma_2
\sum_{a=\pm1}\sum_{b=\pm1}
\delta_{r,a}\delta_{s,b}\\
&- (\gamma_0+ 4\gamma_2)
\sum_{a=\pm1}
(\delta_{r,0}\delta_{s,a} +\delta_{r,a}\delta_{s,0}).
\end{align}
By defining the relative distance $r=x-x'$ and $s=y-y'$ and taking discrete Fourier transform, the source term can be written in Fourier space as
\begin{equation}
\label{Bq-gamma}
B({\bf q}) =
\sum_{\alpha,\beta}
\left(1-e^{iq_\alpha}\right)
\left(1-e^{-iq_\beta}\right)
\Gamma^{\alpha\beta}_{{\bf q}}
\\
\simeq \sum_{\alpha,\beta} q_{\alpha} q_{\beta} \Gamma^{\alpha\beta}_{\bf q},
\end{equation}
where, in the last step, we have used the small-${\bf q}$ approximation.
Now, by substituting the covariance of the fluctuating (noise) current in Fourier space, we explicitly obtain the following expression,
\begin{equation}
B({\bf q}) =
\left[ \gamma_0 + \gamma_2\sum_\alpha\lambda(q_\alpha)
-\gamma_2 \frac{\sum_\alpha\lambda^2(q_\alpha)}
{\sum_\alpha\lambda(q_\alpha)} \right]
\sum_\alpha \lambda(q_\alpha),
\end{equation}
where
\begin{equation}
\lambda(q_\alpha)=2(1-\cos q_\alpha).
\end{equation}
In the steady state, we have $\partial_t C^{mm}_{r,s}=0$, and therefore the static structure factor, which we define as
\begin{equation}
S({\bf q}) =
\sum_{r,s}
C^{mm}_{r,s}\,
e^{i(q_xr+q_ys)},
\end{equation}
can be exactly given by the following expression,
\begin{eqnarray}
\label{Sq-B}
S(q_x,q_y) 
&= \dfrac{B(q_x,q_y)}
{2D\sum_\alpha\lambda(q_\alpha)
- D^{(\rm odd)} \lambda(q_x)\lambda(q_y)}.
\end{eqnarray}
This is the main result of the present study, which we analyze in detail in the subsequent discussions.
We provide heat maps of the static structure factor as a function of wave vector components $q_x$ and $q_y$ and  the equal-time density-density correlation as a function of position vector components $x$ and $y$ in panels (a) and (b) of Fig. \eqref{fig:structure factors}, respectively.
It is interesting to note that, although the odd-diffusion coefficient $D^{\rm (odd)}$
does not contribute to the density relaxation, it enters explicitly into the static structure factor and directly influences the two-point spatial density correlations. 

The off-diagonal components of the diffusion and mobility tensors, which directly encode the chiral coupling between currents along two orthogonal directions, play a crucial role in generating the algebraic correlations observed in the chiral mass-transport processes studied here. Although the underlying mass-transport dynamics is isotropic, these off-diagonal components, $D_{\alpha\beta}$ and $\Gamma^{\alpha\beta}$ ($\alpha\neq\beta$), couple transverse mass currents to density gradients, thereby producing a nontrivial long-wavelength structure in the steady-state correlations of mass and currents. In particular, this chiral coupling generates a $q$-dependent response that survives in the isotropic limit and ultimately gives rise to algebraic spatial decay. The correlations also depend on the diagonal components, $D_{\alpha\alpha}$ and $\Gamma^{\alpha\alpha}$; however, in the absence of the off-diagonal components, the density correlations remain short-ranged (delta-correlated for the present models) and no algebraic correlations are observed.

\subsubsection{Large-distance asymptotics}

To characterize the asymptotic, large-distance behavior of the density correlation function, we consider the small-wavenumber (hydrodynamic) limit, ${\bf q}\to 0$. Expanding the structure factor to the leading orders in the wavenumber, we obtain
\begin{eqnarray}
S(\mathbf q) \simeq
\frac{ \left[
\gamma_0 q^2
-\dfrac{\gamma_0}{12}(q_x^4+q_y^4)
+2\gamma_2 q_x^2q_y^2 \right]
}{ \left[ 
2Dq^2
-\dfrac{D}{6}(q_x^4+q_y^4)
-D^{(\mathrm{odd})}q_x^2q_y^2 \right]
}.
\end{eqnarray}
Expanding the denominator, we rewrite the structure factor in a series $-$ order by order in the wave vector ${\bf q}$,
\begin{align}
\nonumber
S(q_x,q_y) &= S({\bf q} \to 0) + {\cal O}(q^2) + {\cal O}(q^4) 
\\
&=
\frac{\gamma_0}{2D}
+
\left(
\frac{\gamma_2}{D}
+
\frac{\gamma_0D^{(\rm odd)}}{4D^2}
\right)
\frac{q_x^2q_y^2}
{q_x^2+q_y^2}
+
{\cal O}(q^4)
\nonumber\\
&\equiv
s_0 + s_2
\frac{q_x^2q_y^2}
{q_x^2+q_y^2} + {\cal O}(q^4),
\label{Sq-iso}
\end{align}
where we have defined the following density-dependent coefficients,
\begin{equation}
s_0=\frac{\gamma_0}{2D},
\qquad
s_2=
\left( \frac{\gamma_2}{D}
+
\frac{\gamma_0D^{(\rm odd)}}{4D^2} \right).
\label{s0-s2}
\end{equation}
The first term $s_0$ in Eq. \eqref{Sq-iso} is related zero-wave-vector structure factor $S({\bf q} \to 0)$, which is regular around the origin ${\bf q}=0$ and corresponds to a short-range contribution to the correlation function. The second term (involving $s_2$) in Eq. \eqref{Sq-iso} however leads to a nonanalytic contribution proportional to ${q_x^2q_y^2}/{(q_x^2+q_y^2)}$,
which gives rise to power-law correlations in the systems.
Importantly, the above analysis in Eqs. \eqref{Sq-iso} and \eqref{s0-s2} shows that  the amplitude of the asymptotic power-law decay of the density correlations is in fact determined by the {\it off-diagonal} components of both the diffusion and mobility tensors. Furthermore, as discussed in Sec. \eqref{sec-cusp}, the mobility coefficients $\gamma_0$ and $\gamma_2$ themselves depend on the odd-diffusion strength $D^{\rm (odd)}$ in a nontrivial way.
Note that, using Eqs \eqref{Bq-gamma} into \eqref{Sq-B} and then after doing some straightforward algebraic manipulations, we can derive the following identity,
\begin{align}
    S({\bf q} \to 0) \equiv s_0 = \lim_{{\bf q} \to 0} \frac{ \sum_{\alpha \beta} \chi^{\alpha \beta} q_{\alpha}q_{\beta} }{\sum_{\alpha \beta} D_{\alpha \beta} q_{\alpha}q_{\beta} },
\end{align}
where the zero-wavenumber structure factor, or the scaled mass fluctuation (``compressibility''), is expressed in terms of the diffusion and mobility tensor.

\begin{figure*}
\centering

\begin{minipage}{0.485\linewidth}
\centering
\begin{overpic}[width=\linewidth]{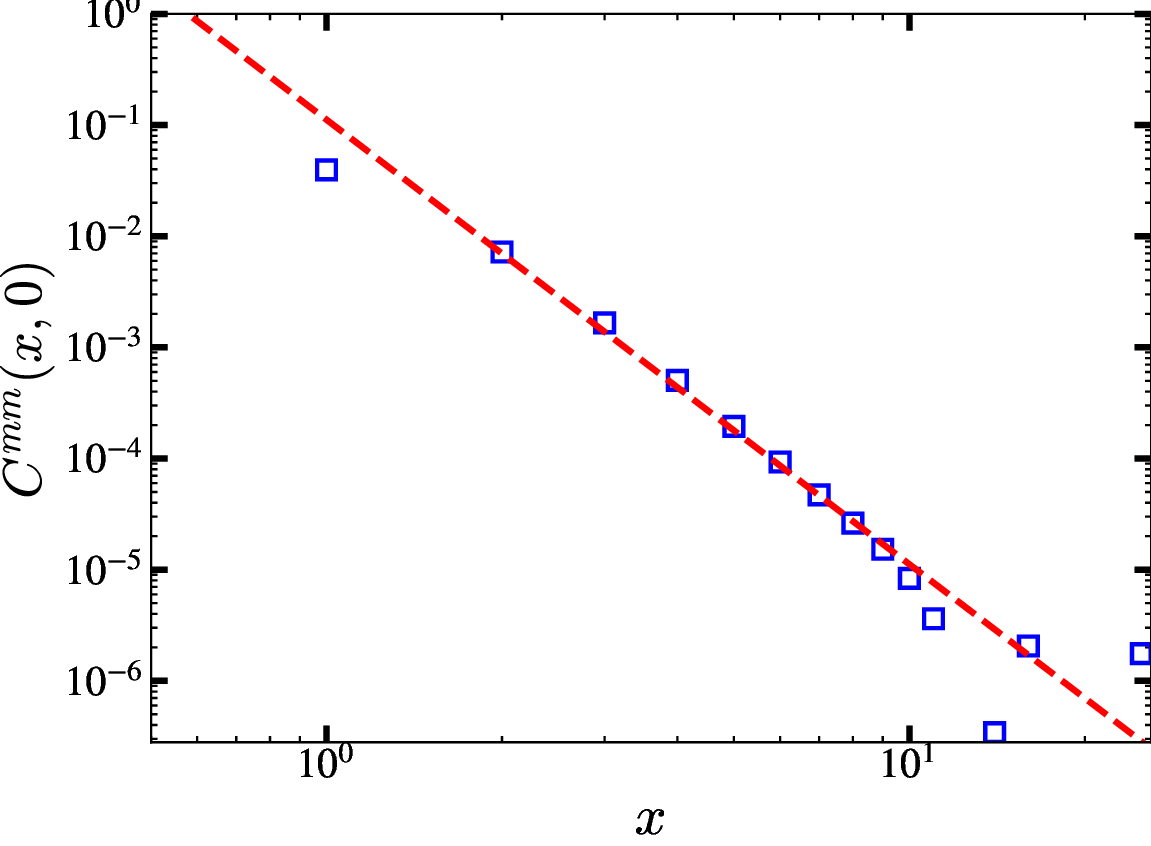}
\put(4,77){\large\bfseries (a)}
\end{overpic}
\end{minipage}
\hfill
\begin{minipage}{0.485\linewidth}
\centering
\begin{overpic}[width=\linewidth]{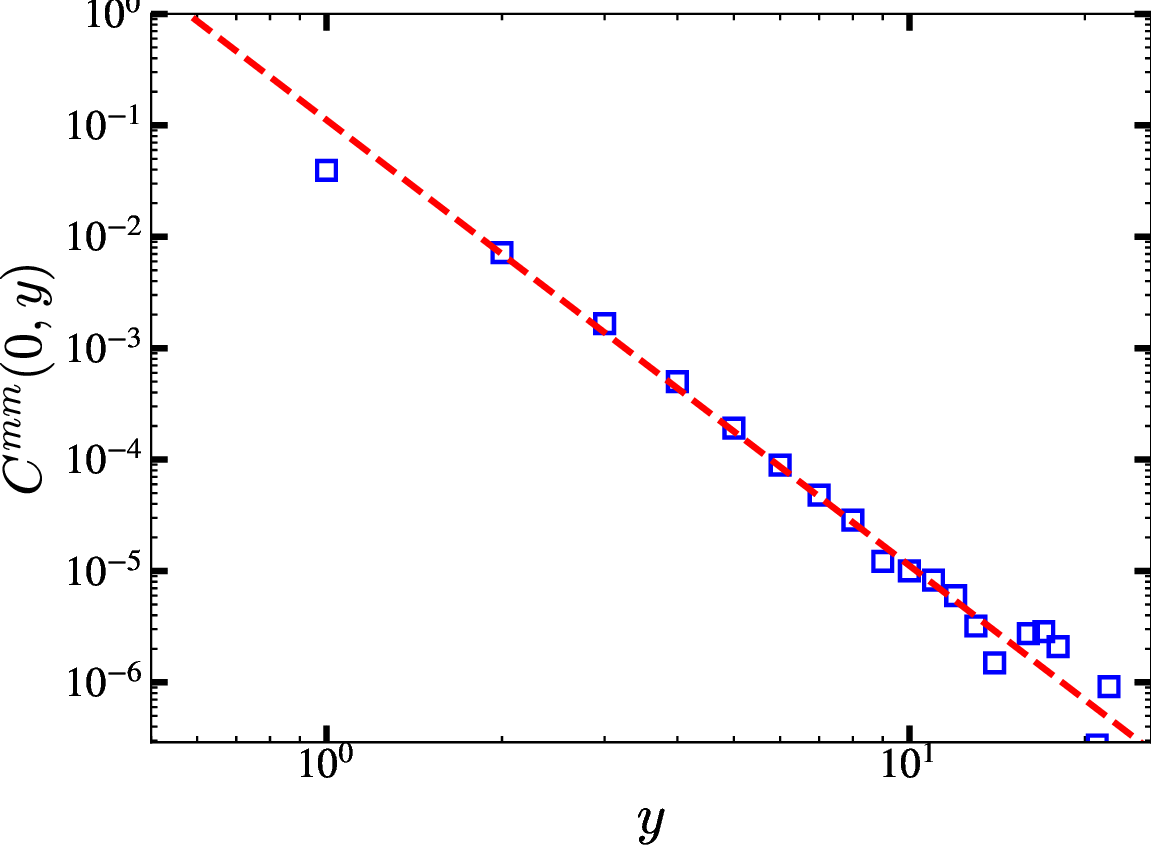}
\put(4,77){\large\bfseries (b)}
\end{overpic}
\end{minipage}

\caption{\textit{Power-law two-point spatial density correlations in the isotropic chiral mass-transport model in the steady state.}
Equal-time density-density correlation function measured along (a) the $x$ direction and (b) the $y$ direction in the nonequilibrium steady state. Symbols denote Monte Carlo simulation results, while solid lines show the exact (asymptotic) theoretical prediction. The two-point density correlations exhibit the characteristic algebraic decay  $C^{mm}(x,y)\sim |{\bf r}|^{-4}$ with distance $|{\bf r}|=\sqrt{x^2+y^2} \gg 1$; see Eq.~\eqref{Cr4}. The simulation parameters are as follows: system size $L_x\times L_y=800\times800$, global density $\bar \rho=1$, and mass-chipping (retention) parameters $\lambda_1=0.4$ and $\lambda_2=0.6$.  }

\label{fig:correlation for isotropy chiral}
\end{figure*}

The nonanalytic contribution to the structure factor gives rise to the asymptotic real-space density correlation at large distance $|{\bf r}| \gg 1$ with position vector denoted as ${\bf r} = (x,y)$,
\begin{align}\label{eq:Cmmxy_assym}
C^{mm}(x,y) &\simeq
\frac{s_2}{(2\pi)^2}
\int_{-\infty}^{\infty} dq_x dq_y
\frac{q_x^2 q_y^2}{q_x^2+q_y^2}
e^{-i(q_x x + q_y y)}\notag\\
&=
s_2 \partial_x^2 \partial_y^2
\frac{1}{(2\pi)^2}
\int_{-\infty}^{\infty}
dq_x dq_y
\frac{e^{-i{\bf q}\cdot{\bf r}}}{q_x^2+q_y^2}\notag\\
&=
s_2
\frac{3(6x^2y^2 - x^4 - y^4)}
{\pi (x^2+y^2)^4}.
\end{align}
Now, expressing the correlation function in polar coordinates,
$(
x=|{\bf r}| \cos\theta,
y=|{\bf r}| \sin\theta
)$
we obtain, for large $|{\bf r}|$, the density correlation function
\begin{align}
\label{Cr4}
C^{mm}(|{\bf r}|,\theta) \simeq
-\frac{3s_2}{\pi} \frac{\cos(4 \theta)}{|{\bf r}|^4}.
\end{align}
Thus, the density correlation decays algebraically as
$C^{mm}(|{\bf r}|,\theta)\sim |{\bf r}|^{-4},$
demonstrating a generic algebraic decay of two-point density correlations in the steady state of a system with  {\it chiral, yet isotropic}, hopping of masses.
In Fig. \ref{fig:correlation for isotropy chiral}, we plot the density correlation along the $x$-direction (a) and $y$-direction (b) for a system of size $L_x\times L_y=800\times800$, with global density $\rho=1$ and chipping parameters $\lambda_1=0.4$ and $\lambda_2=0.6$. The asymptotic decay of the density correlation in both directions shows good agreement with the theoretical prediction given in Eq. \eqref{Cr4}.

It is well known that anisotropic hopping can give rise to power-law decay of density correlations in systems with mass-conserving dynamics \cite{Grinstein1990Apr, Garrido1990Aug}. The origin of this anisotropic power law can be understood through an electrostatic analogy, in which the correlation function satisfies a Poisson equation with a quadrupolar charge distribution represented by a second-rank tensor, resulting in a $1/r^2$ decay in two dimensions. In the present case, however, isotropy symmetry eliminates the lower-order anisotropic contribution, and the corresponding effective charge distribution having $4$th-order multipole moment is instead associated with a rank-four tensor, leading to the faster $1/|{\bf r}|^4$ decay.
The angular dependence of $\cos(4 \theta)$ in Eq. \eqref{Cr4}
further reveals the characteristic fourfold structure of the correlations, with alternating positive and negative lobes separated by nodal directions along which the correlation vanishes.
Notably, the mechanism proposed in the present work is distinct from the anisotropic mechanism discussed in Ref. \cite{Garrido1990Aug}. In the present case, the simultaneous transfer of mass across multiple lattice bonds is crucial for implementing chirality in the dynamics and, in turn, gives rise to the generic algebraic decay described by Eq. \eqref{Cr4}.

The unknown coefficients, $M_0$ and $M_1$  are determined self-consistently from the structure factor by using the following two conditions:
\begin{align}
\label{cond1}
M_0
&=
\frac{1}{(2\pi)^2}
\int_{\rm BZ}
S({\bf q}|M_0, M_1)\,
d^2{\bf q}
+\rho^2,
\\ 
\label{cond2}
M_1
&=
\frac{1}{(2\pi)^2}
\int_{\rm BZ}
e^{-iq_x}
S({\bf q}|M_0, M_1)\,
d^2{\bf q}
+\rho^2.
\end{align}
Solving these coupled equations yields the exact onsite and nearest-neighbor mass correlations. Consequently, the coefficients $\gamma_0$ and $\gamma_2$ [see Eqs. \eqref{g0} and \eqref{g2}] appearing in the expression of the mobility tensor, as well as $s_0$ and $s_2$ in the expression of the static structure factor, are fully determined in terms of the global density and the other (microscopic) parameters of the model. In practice, however, obtaining the exact closed-form expressions for $M_0$ and $M_1$, and hence for $\gamma_0$ and $\gamma_2$, is rather cumbersome. We therefore determine these coefficients next by examining the leading-order behavior of the structure factor in Eq. \eqref{Sq-iso}.

\subsubsection{Density fluctuation and cusp singularity}
\label{sec-cusp}

Here we discuss an interesting consequence of odd diffusion, which has a dual role in governing density fluctuations. At moderate chirality, odd transport promotes circulating mass flow and suppresses long-wavelength density fluctuations, leading to an optimal chirality at which the zero-wavenumber structure factor $s_0$ is minimized. Upon further increasing the odd-diffusion strength toward its maximal admissible value, however, this suppression gives way to a singular response: the mobility and mass fluctuations develop a square-root, cusp-like singularity. To elucidate this behavior, we first analyze the mobility coefficient $\gamma_0$, which depends on $M_0$ and $M_1$.

As discussed in the previous section, the parameters $M_0$ and $M_1$ can be calculated by solving the following coupled equations,
\begin{align}
    M_0 - \rho^2  = \frac{\gamma_0(M_0, M_1)}{2D}, \\
    M_1 - \rho^2  = - \frac{3 s_2(M_0, M_1)}{\pi}.
\end{align}
In the following analysis, we employ the leading-order form of the structure factor, given in Eq. \eqref{Sq-iso}, to obtain explicit expressions for the coefficients $M_0$ and $M_1$ in terms of the density and the model parameters.
For convenience, we define the coefficients
\begin{align}
A_1
&=
\frac{1}{2}\mu_2
\Big(
\tilde{\lambda}_1^2
+\tilde{\lambda}_2^2
+\mu_2\tilde{\lambda}_1^2\tilde{\lambda}_2^2
\Big),
\notag\\
A_2
&=
\mu_1\mu_2
\tilde{\lambda}_1\tilde{\lambda}_2^2,
\notag\\
A_3 
&=
-\frac{1}{4}\mu_1^2
\tilde{\lambda}_1\tilde{\lambda}_2 \nonumber,
\\
A_4
&=
-\frac{1}{4}\mu_1\mu_2
\tilde{\lambda}_1^2\tilde{\lambda}_2 \nonumber.
\end{align}
By substituting the above coefficients into the following relation,
\begin{align}
M_0-\rho^2=\frac{\gamma_0}{2D},
\end{align}
we obtain
\begin{align}
(2D - A_1)M_0 - A_2 M_1
= 2D \rho^2.
\label{eq:M0eq}
\end{align}
Similarly, by using the relation
\begin{align}
M_1-\rho^2
=
-\frac{3}{\pi}
\left(
\frac{\gamma_2}{D}
+
\frac{\gamma_0D^{(\rm odd)}}{4D^2}
\right),
\end{align}
we obtain the following equation,
\begin{align}
M_1-\rho^2
=
-\frac{3}{\pi}
\Bigg[
\left(
\frac{A_4}{D}
+
\frac{A_1 D^{(\rm odd)}}{4D^2}
\right)M_0
\nonumber \\ 
+
\left(
\frac{A_3}{D}
+
\frac{A_2 D^{(\rm odd)}}{4D^2}
\right)M_1
\Bigg].
\end{align}
Now, solving the above set of equations, we obtain the desired solution for $M_0$ and $M_1$ as given below:
\begin{align}
M_0 =
\rho^2
\frac{
2D
\Big[
\pi D^2
+
3\left(
A_3 D+\frac{A_2 D^{(\rm odd)}}{4}
\right)
\Big]
+
\pi D^2 A_2
}
{\Delta},
\\
M_1
=
\rho^2
\frac{
\pi D^2(2D - A_1)
-
6D
\left(
A_4 D+\frac{A_1 D^{(\rm odd)}}{4}
\right)
}
{\Delta},
\end{align}
which can be explicitly written as a function $D$ and $D^{(\rm odd)}$ by inverting the parameters $\tilde \lambda_1$ and $\tilde \lambda_2$,
\begin{align}
\tilde{\lambda}_{1}
&=
\frac{1}{\mu_1}
\left[
2D-D^{(\rm odd)}
+
\sqrt{
\left(2D-D^{(\rm odd)}\right)^2
-2D^{(\rm odd)}
}
\right],
\\
\tilde{\lambda}_{2}
&=
\frac{1}{\mu_1}
\left[
2D-D^{(\rm odd)}
-
\sqrt{
\left(2D-D^{(\rm odd)}\right)^2
-2D^{(\rm odd)}
}
\right].
\end{align}
Thus, the coefficients $A_i$'s, and consequently $M_0$ and $M_1$, can be expressed explicitly in terms of $D$ and $D^{(\rm odd)}$.
Finally, the mobility coefficients are given by
\begin{align}
\gamma_0(D, D^{(\rm odd)}) &= A_1 M_0 + A_2 M_1,
\\
\gamma_2(D, D^{(\rm odd)}) &= A_3 M_1 + A_4 M_0.
\end{align}
These expressions yield, in a closed form, the mobility coefficients and hence the two-point spatial density correlation function entirely in terms of the even and odd diffusion coefficients, $D$ and $D^{(\rm odd)}$. The resulting expressions are still algebraically quite cumbersome and do not provide additional physical insight in their full form; we therefore refrain from displaying them explicitly. Instead, we focus on their qualitative behavior when the odd-diffusion strength varies around zero or approaches its maximal possible value.

From the above analysis, we find that the system crosses over continuously from a completely local structure factor at $D^{\rm odd}=0$ (achiral) to one with a nonanalytic angular dependence and algebraic $1/r^4$ correlations as soon as odd diffusion (chirality) is switched on, keeping normal diffusion coefficient constant. Indeed, the onset of chirality immediately induces power-law correlations.
More specifically, the small-$D^{\rm odd}$ expansion has the following physical implications:
At $D^{\rm odd}=0$, there are no power-law density correlations as only one microscopic mode of transport—that of the normal diffusion—is present ($\tilde \lambda_2=0$ and thus no odd diffusion). At small $D^{\rm odd}$, the structure factor is analytic,
\begin{align}
    \gamma_0 &= \gamma_0^{(0)} +  \gamma_0^{(1)} D^{\rm (odd)} + {\cal O}((D^{\rm (odd)})^2),
    \label{gamma0-d-odd}\\
    \gamma_2 &= \gamma_2^{(1)} D^{\rm (odd)} + {\cal O}((D^{\rm (odd)})^2),
    \\
    s_2 &= \left( \frac{\gamma_2}{D} + \frac{\gamma_0 D^{\mathrm{(odd)}}}{4D^2} \right) \simeq \left( \frac{\gamma_2^{(1)}}{D} + \frac{\gamma_0^{(0)} }{4D^2} \right) D^{\mathrm{(odd)}},
\end{align}
where $\gamma_k^{(l)}$'s denote constant coefficients in the corresponding power-series expansions.
Note that, as the prefactor $\gamma_0^{(1)} < 0$ is negative, the mobility coefficient $\gamma_0$ in Eq. \eqref{gamma0-d-odd} in fact decreases with increasing $D^{\rm (odd)}$, consistent with the suppression of density fluctuations by the enhanced circular orbital motion of the masses.
However, upon further increasing the chirality, this perturbative behavior changes. Quite remarkably, there exists an optimal odd diffusion at which the mobility, and therefore density fluctuations, are minimized and then they start increasing until they reach the maximally admissible endpoint, which corresponds to a branch-point singularity of the underlying transport modes. Indeed, expanding around the branch-point, the mobility acquires a nonanalytic form,
\begin{align}
    \gamma_0(D^{\rm (odd)}) = \gamma_c + a_1 \sqrt{(D_c^{\rm (odd)}-D^{\rm (odd)})},
\end{align}
where $\gamma_{0,c}$ denotes the limiting value at the branch point and  $a_1>0$ is a model-dependent coefficient.
That is, around the branch-point, turning on a small odd diffusion activates the second mode of transport with strength $\tilde \lambda_2 \propto D^{\rm (odd)}$. This generates $\gamma_2 \propto D^{\rm (odd)}$, so the amplitude of the nonanalytic part of the structure factor—and therefore the $1/r^4$ power-law correlation—grows linearly with $D^{\rm (odd)}$ at leading order, $s_2 \propto D^{\rm (odd)}$.
So the branch-point singularity appears only when the two modes of transport (odd and normal diffusion) become comparable in magnitude; it is not manifest in the small-$D^{\rm (odd)}$ expansion though. Thus, the weak-chirality regime is governed by regular perturbation theory, whereas the endpoint is governed by the square-root (cusp) singularity of the mobility as a function of $(D^{\rm (odd)}_c - D^{\rm (odd)})$; see Fig. \ref{fig:cusp}.

\begin{figure}
    \centering
    \includegraphics[width=0.9\linewidth]{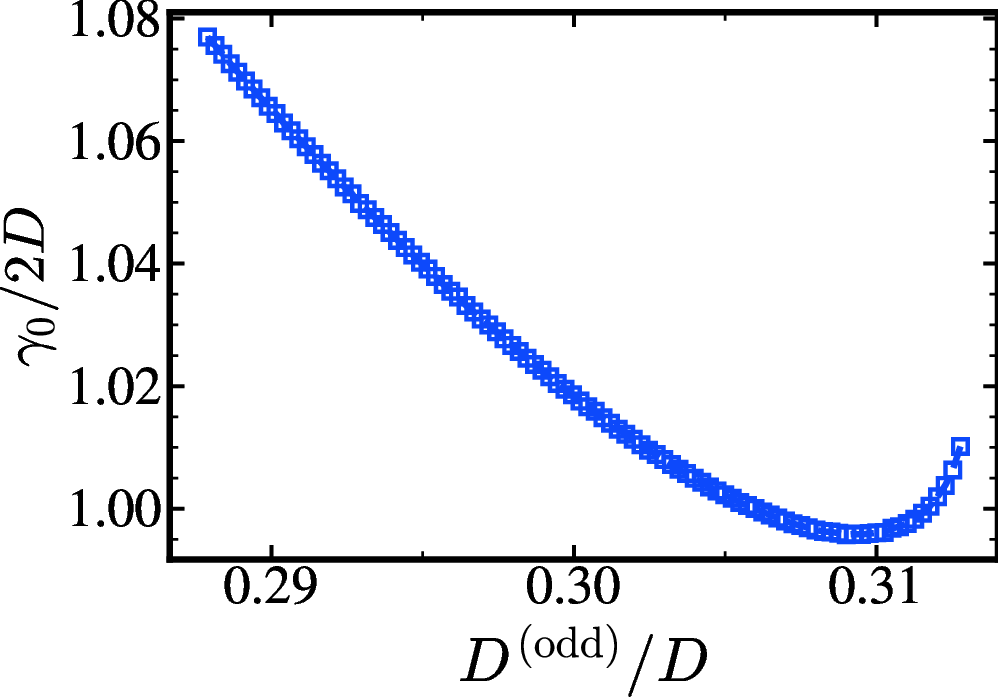}
    \put(-230, 150){\large\bfseries (a)}
    \\
     \includegraphics[width=0.9\linewidth]{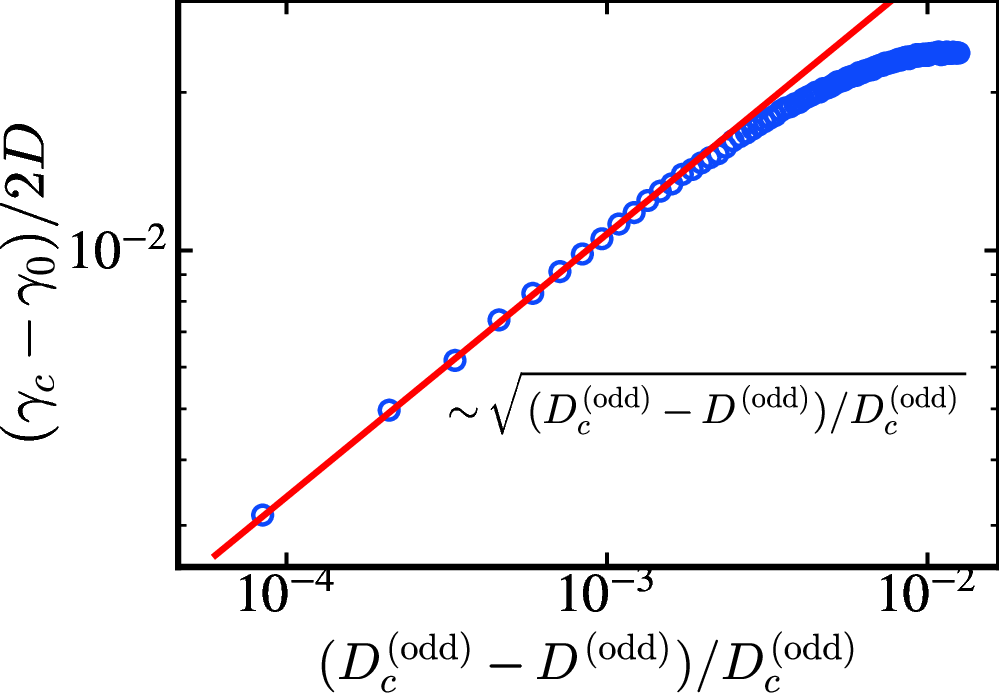}
    \put(-230, 150){\large\bfseries (b)}
    \caption{ {\it Density fluctuations and the mobility near the branch-cut point.} Panel (a): The zero-wavenumber structure factor $S({\bf q} \to 0) \equiv s_0 = \gamma_0/2D$ obtained form simulations is plotted as a function of the scaled odd-diffusion coefficient $D^{\rm odd}/D$ near the branch-cut point $D^{\rm(odd)}=D^{\rm(odd)}_c$.  Panel (b): The deviation $(s_0^{\rm (c)}-s_0)=(\gamma_c-\gamma_0)/2D$, where $s_o^{\rm (c)}=s_0(D^{\rm(odd)}_c)=\gamma_c/2D$ and $\gamma_c=\gamma_0(D^{\rm(odd)}_c)$, is plotted as a function of $(D^{\rm(odd)}_c-D^{\rm (odd)})/D^{\rm(odd)}_c$ around the branch-cut point $D^{\rm(odd)}_c$; red (solid) line is a guide to the eye. The normal bulk-diffusion coefficient is kept fixed.  }
    \label{fig:cusp}
\end{figure}

\subsection{A fluctuation--dissipation relation}\label{subsec:FDR}


The fluctuation–dissipation relation (FDR), which connects transport coefficients (obtained through a linear response theory) to steady-state fluctuations, is a cornerstone of equilibrium statistical mechanics. A prototypical example is the Einstein relation (ER) \cite{Einstein1905Jan}, which relates the diffusion coefficient to the mobility and the strength of equilibrium fluctuations. Closely related relations underlie the Johnson–Nyquist formula \cite{Johnson1928Jul, Nyquist1928Jul} and the Green–Kubo expressions for transport coefficients \cite{Kubo1957Jun, Green1954Mar}, establishing a general connection between dissipation, response, and spontaneous equilibrium fluctuations. In equilibrium, such relations are consequences of microscopic time-reversal symmetry (or detailed balance) together with the Boltzmann–Gibbs form of the steady-state distribution.

This relationship is, however, not generally maintained in nonequilibrium steady states, where detailed balance is broken and probability currents persist in the steady state. Consequently, there is no universal principle that fixes the relation between transport coefficients, response functions, and steady-state fluctuations. Einstein-type relations \cite{Blickle2007May} and generalized fluctuation–dissipation relations \cite{Maes2020Mar} have nevertheless been derived for several classes of diffusive nonequilibrium systems, including systems described by fluctuating hydrodynamics and macroscopic fluctuation theory \cite{Bertini2001Jul,  Bertini2015Jun}. However, the precise conditions under which such relations hold, as well as the manner in which they are modified by nonequilibrium driving, remain only partially understood.

Indeed, identifying the conditions under which transport coefficients can be expressed in terms of steady-state fluctuations remains an important question in nonequilibrium statistical physics. For an equilibrium system in one dimension, the fluctuation–dissipation relation connects the mobility $\chi$, bulk-diffusion coefficient $D$, and the space-integrated static density–density correlation function according to the Einstein relation,
\begin{align}
\chi = D \int_{-\infty}^{\infty} dx ,
C^{mm}(x) = D S({\bf q}\to{\bf 0}).
\label{eq}
\end{align}
In two or higher dimensions, the transport coefficients are, in general, tensors, and the corresponding FDR can be written in a matrix form as
\begin{align}
\boldsymbol{\chi} = \mathbb{D}\mathbb{S},
\label{FDR-iso}
\end{align}
where $\boldsymbol{\chi}$ and $\mathbb{D}$ denote the mobility and diffusion tensors, respectively. For an isotropic system with a single conserved field (single-species), structure-factor matrix in the the zero-wavenumber limit is simply proportional to the identity matrix,
\begin{equation}
\mathbb{S}=s_0
\begin{pmatrix}
1 & 0\\
0 & 1
\end{pmatrix},
\end{equation}
with $s_0=\lim_{{\bf q}\to{\bf 0}}S({\bf q})$. For equilibrium systems with short-ranged interactions, the limits $q_x\to0$ and $q_y\to0$ usually commute, so that $s_0$ is independent of the direction in which the zero-wave-vector limit is taken.
Equation \eqref{FDR-iso} expresses a fundamental connection between macroscopic transport and steady-state fluctuations. This equilibrium-like relation, however, does not hold for the chiral mass-transport processes considered here when the full diffusion tensor is used. In particular,
\begin{align}
\boldsymbol{\chi}
&= \frac{1}{2}
\begin{pmatrix}
\gamma_0(\bar{\rho}) & 0\\
0 & \gamma_0(\bar{\rho})
\end{pmatrix}
\neq
\begin{pmatrix}
D & -D^{\rm odd}\\
D^{\rm odd} & D
\end{pmatrix}
\begin{pmatrix}
s_0 & 0\\
0 & s_0
\end{pmatrix},
\end{align}
or, equivalently, $\boldsymbol{\chi}\neq\mathbb{D}\mathbb{S}$. The violation arises from the antisymmetric, or odd, off-diagonal components of the diffusion tensor, which are a direct consequence of the chiral dynamics.
Nevertheless, decomposing the diffusion tensor into its symmetric and antisymmetric parts,
\begin{align}
\mathbb{D}=\mathbb{D}_S+\mathbb{D}_A,\qquad
\mathbb{D}_S=\frac{\mathbb{D}+\mathbb{D}^{T}}{2},\qquad
\mathbb{D}_A=\frac{\mathbb{D}-\mathbb{D}^{T}}{2},
\end{align}
reveals a generalized fluctuation–dissipation relation involving only the dissipative, symmetric part of the transport tensor:
\begin{align}
\boldsymbol{\chi}=\mathbb{D}_S\mathbb{S}.
\end{align}
Thus, while chirality modifies the conventional equilibrium-like relation involving the full diffusion tensor, the mobility remains related to steady-state fluctuations through the symmetric part of the diffusion tensor. This relation provides a natural nonequilibrium generalization of the equilibrium FDR in Eq. \eqref{FDR-iso}.

\section{Summary and concluding remarks}\label{sec:summary}

In this paper, we have investigated nonequilibrium transport processes with mass-conserving dynamics that break both time-reversal and mirror (parity) symmetries. To this end, we have introduced a class of interacting Markov jump processes on a square lattice and constructed minimal models in which chirality emerges from the simultaneous, coordinated hopping of masses across multiple spatially separated bonds. Although the individual hopping events are {\it isotropic}, their coordinated dynamics acquires a preferred sense of rotation, thereby giving rise to {\it chiral} transport.

We consider both periodic and open boundary conditions; in both cases, total mass remains conserved.
On a periodic domain, the dynamics is purely diffusive, and the bulk current in the  steady state vanishes. In contrast, in the presence of open (reflecting) boundaries, or “walls,” the system supports a {\it nonzero} steady-state {\it edge current} flowing tangential to the wall. Despite exhibiting nontrivial many-body correlations, these models remain exactly solvable, allowing us to derive the transport coefficients (density-dependent in general) directly from the underlying microscopic dynamics and to develop a microscopic theory of large-scale fluctuations in interacting chiral systems driven out of equilibrium.

We demonstrate that odd diffusion generically provides an alternative and robust mechanism for the emergence of scale-invariant (algebraic) two-point spatial density correlations in nonequilibrium steady state. Notably, this mechanism is qualitatively distinct from the well-known mechanism based on anisotropic hopping proposed in Refs. \cite{Grinstein1990Apr, Garrido1990Aug, Maes1990Nov}, as it is applicable even to systems with isotropic hopping rates.
In particular, we exactly calculate the equal-time density-density correlation function in the steady state and show that the correlation function exhibits a universal algebraic decay, $C({\bf r}) \sim |{\bf r}|^{-4}$, at large distances $|{\bf r}| \gg 1$. 
Remarkably, the amplitude of the power-law correlations gets contributions from both the parity-breaking diffusion coefficient $D^{(\rm odd)}$ and the strength $\gamma_2$ of the orthogonal ``noise''-current correlations (equivalently, the mobility tensor). Thus, these algebraic correlations are  determined jointly by the off-diagonal (odd) components of the diffusion tensor and the off-diagonal components of the mobility tensor.
By determining the structure factor, we further established a nonequilibrium Green–Kubo-like fluctuation–dissipation relation that expresses the steady-state density fluctuations entirely in terms of the bulk diffusion and mobility tensors.

Another interesting consequence of odd diffusion is its nontrivial influence on fluctuations in the sense that the odd diffusion plays a dual role. At moderate values, it suppresses long-wavelength density fluctuations by promoting circulating transport, leading to a minimum ``compressibility'' (structure factor in the zero-wavenumber limit). As the maximal admissible strength of odd diffusion, $D_c^{({\rm odd })}$, is approached, however, it produces a square-root (cusp) nonanalytic response of the mobility and the mass fluctuation. Thus, chirality first stabilizes and then renders the system increasingly sensitive to the change of odd-diffusion strength. There exists an optimal chirality at which density fluctuations are maximally suppressed. That is, increasing chiral flow first enhances mixing, density inhomogeneities are reduced, and ``compressibility'' decreases. However, upon further increasing the chirality, the compressibility and the mobility start increasing sharply and become a singular function of the chirality parameter. 
This behavior is somewhat reminiscent of caustic-like catastrophes encountered in other areas of physics, where optical or acoustic intensity becomes concentrated along singular loci \cite{berry1980catastrophe,Berry1982Aug, Kravtsov1983Dec}.

Indeed, the present work opens several directions for future investigation. An important question is the extent to which the mechanism identified here persists beyond the exactly solvable models considered here, including systems with more general interactions or active driving. It would also be interesting to understand how odd diffusion affects other mechanisms known to generate algebraic correlations, such as anisotropic hopping, additional conservation laws, and external driving. Another natural direction is to explore the role of odd transport in systems with open boundaries \cite{Dolai2025Oct}, where the interplay between chirality, boundary-induced currents, and fluctuations may give rise to qualitatively new nonequilibrium phenomena. 
More broadly, our results underscore the role of odd transport coefficients as fundamental ingredients governing the large-scale organization of fluctuations in driven many-body systems, rather than merely as signatures of nondissipative transport.

\section{Acknowledgement}

We thank Deepak Dhar for useful discussions and reading of the manuscript.

\appendix

\section{Infinitesimal update rules for mass and current}

Here we provide the infinitesimal-time update rules for local mass variables for the general version of the chiral mass transport process:
\begin{widetext}
\begin{align}
\label{eq:mass_update}
m(x,y,t+dt)=
    \begin{cases}
        \textbf{event}         & \textbf{prob.} \\
        m(x,y,t) - \tilde{\xi}_1 \tilde{\lambda}_x m(x,y,t) & p_1 dt\\
        m(x,y,t) - \tilde{\xi}_1 \tilde{\lambda}_x m(x,y,t) & p_2 dt\\
        m(x,y,t) - \tilde{\xi}_1 \tilde{\lambda}_y m(x,y,t) &  p_3 dt\\
        m(x,y,t) - \tilde{\xi}_1 \tilde{\lambda}_y m(x,y,t) &  p_4 dt\\
        m(x,y,t) + [m(x,y-1,t) + \tilde{\xi}_1\tilde{\lambda}_x m(x-1,y-1,t)]\tilde{\xi}_2 \tilde{\lambda}_{xy} & p_1 dt\\
        m(x,t,t) + [m(x+1,y,t) + \tilde{\xi}_1\tilde{\lambda}_y m(x+1,y-1,t)]\tilde{\xi}_2 \tilde{\lambda}_{yx} & p_3 dt\\
        m(x,y,t) + [m(x,y+1,t) + \tilde{\xi}_1 \tilde{\lambda}_x m(x+1,y+1,t)]\tilde{\xi}_2 \tilde{\lambda}_{xy} & p_2 dt\\
        m(x,y,t) + [m(x-1,y,t) + \tilde{\xi}_1 \tilde{\lambda}_y m(x-1,y+1,t)]\tilde{\xi}_2 \tilde{\lambda}_{yx} & p_4 dt\\
        m(x,y,t) + \tilde{\xi}_1 \tilde{\lambda}_x m(x+1,y,t) - [m(x,y,t) + \tilde{\xi}_1 \tilde{\lambda}_x m(x+1,y,t)]\tilde{\xi}_2 \tilde{\lambda}_{xy} & p_2 dt\\
        m(x,y,t) + \tilde{\xi}_1 \tilde{\lambda}_x m(x-1,y,t) - [m(x,y,t) + \tilde{\xi}_1 \tilde{\lambda}_x m(x-1,y,t)]\tilde{\xi}_2 \tilde{\lambda}_{xy} & p_1 dt\\
        m(x,y,t) + \tilde{\xi}_1 \tilde{\lambda}_y m(x,y+1,t) - [m(x,y,t) + \tilde{\xi}_1 \tilde{\lambda}_y m(x,y+1,t)]\tilde{\xi}_2 \tilde{\lambda}_{yx} & p_4 dt\\
        m(x,y,t) + \tilde{\xi}_1 \tilde{\lambda}_y m(x,y-1,t) - [m(x,y,t) + \tilde{\xi}_1 \tilde{\lambda}_y m(x,y-1,t)]\tilde{\xi}_2 \tilde{\lambda}_{yx} & p_3 dt\\
        m(x,y,t)  &1-\sum {\rm (prob.)}
        \end{cases}
 \end{align}
 \end{widetext}
These update rules can be used to derive the time-evolution equation \eqref{den-cor} for equal-time correlation function for masses at two spatial points.

Similarly, the infinitesimal-time update rules for bond currents are provided below for the current along $x$-direction,
\begin{widetext}
\begin{align}
\label{curr-update1}
Q_x(x,y,t+dt)=
    \begin{cases}
        \textbf{event}         & \textbf{prob.} \\
        Q_x (x,y,t) + \tilde{\xi}_1 \tilde{\lambda}_x m(x,y,t) & p_1 dt\\
        Q_x (x,y,t) - [m(x+1,y,t) + \tilde{\xi}_1 \tilde{\lambda}_y m(x+1,y-1,t)]\tilde{\xi}_2 \tilde{\lambda}_{yx} & p_3 dt \\ 
        Q_x (x,y,t) + [m(x,y,t) + \tilde{\xi}_1 \tilde{\lambda}_y m(x,y+1,t)]\tilde{\xi}_2 \tilde{\lambda}_{yx} & p_4 dt \\
        Q_x (x,y,t) - \tilde{\xi}_1 \tilde{\lambda}_x m(x+1,y,t) & p_2 dt\\
        Q_x (x,y,t) & 1-\sum (all \ prob.)\\
        \end{cases}
\end{align}
and that along $y$-direction,
 \begin{align}
 \label{curr-update2}
Q_y(x,y,t+dt)=
    \begin{cases}
        \textbf{event}         & \textbf{prob.} \\
        Q_y (x,y,t) + \tilde{\xi}_1 \tilde{\lambda}_y m(x,y,t) & p_3 dt\\
        Q_y (x,y,t) -  [m(x,y+1,t) + \tilde{\xi}_1 \tilde{\lambda}_x m(x+1,y+1,t)]\tilde{\xi}_2 \tilde{\lambda}_{xy} & p_2 dt \\ 
        Q_y (x,y,t) + [m(x,y,t) + \tilde{\xi}_1 \tilde{\lambda}_x m(x-1,y,t)]\tilde{\xi}_2 \tilde{\lambda}_{xy} & p_1 dt \\
        Q_y (x,y,t) - \tilde{\xi}_1 \tilde{\lambda}_y m(x,y+1,t) & p_4 dt\\
        Q_y (x,y,t) & 1-\sum (all \ prob.)\\
        \end{cases}
\end{align}
\end{widetext}
These update rules can be used to obtain the time-evolution equation \eqref{current-cor} for equal-time correlation function for currents at two spatial points and the associated mobility tensor.

\bibliographystyle{apsrev4-2}
\bibliography{ref}

@article{Aldous1995Jun,
	author = {Aldous, D. and Diaconis, P.},
	title = {{Hammersley's interacting particle process and longest increasing subsequences}},
	journal = {Probab. Theory Related Fields},
	volume = {103},
	number = {2},
	pages = {199--213},
	year = {1995},
	month = jun,
	issn = {1432-2064},
	publisher = {Springer-Verlag},
	doi = {10.1007/BF01204214}
}

@article{Kipnis1982Jan,
	author = {Kipnis, C. and Marchioro, C. and Presutti, E.},
	title = {{Heat flow in an exactly solvable model}},
	journal = {J. Stat. Phys.},
	volume = {27},
	number = {1},
	pages = {65--74},
	year = {1982},
	month = jan,
	issn = {1572-9613},
	publisher = {Kluwer Academic Publishers-Plenum Publishers},
	doi = {10.1007/BF01011740}
}

@article{Bertini2001Jul,
	author = {Bertini, L. and De Sole, A. and Gabrielli, D. and Jona-Lasinio, G. and Landim, C.},
	title = {{Fluctuations in Stationary Nonequilibrium States of Irreversible Processes}},
	journal = {Phys. Rev. Lett.},
	volume = {87},
	number = {4},
	pages = {040601},
	year = {2001},
	month = jul,
	issn = {1079-7114},
	publisher = {American Physical Society},
	doi = {10.1103/PhysRevLett.87.040601}
}

@article{Bertini2015Jun,
	author = {Bertini, Lorenzo and De Sole, Alberto and Gabrielli, Davide and Jona-Lasinio, Giovanni and Landim, Claudio},
	title = {{Macroscopic fluctuation theory}},
	journal = {Rev. Mod. Phys.},
	volume = {87},
	number = {2},
	pages = {593--636},
	year = {2015},
	month = jun,
	issn = {1539-0756},
	publisher = {American Physical Society},
	doi = {10.1103/RevModPhys.87.593}
}

@article{Chen2024Mar,
	author = {Chen, Letian and Huynh, Hoai Nguyen and Pruessner, Gunnar},
	title = {{Correlation, crossover, and broken scaling in the Abelian Manna model}},
	journal = {Phys. Rev. Res.},
	volume = {6},
	number = {1},
	pages = {013248},
	year = {2024},
	month = mar,
	publisher = {American Physical Society},
	doi = {10.1103/PhysRevResearch.6.013248}
}

@article{Grinstein1990Apr,
	author = {Grinstein, G. and Lee, D.-H. and Sachdev, Subir},
	title = {{Conservation laws, anisotropy, and {\textasciigrave}{\textasciigrave}self-organized criticality'' in noisy nonequilibrium systems}},
	journal = {Phys. Rev. Lett.},
	volume = {64},
	number = {16},
	pages = {1927--1930},
	year = {1990},
	month = apr,
	publisher = {American Physical Society},
	doi = {10.1103/PhysRevLett.64.1927}
}

@article{Garrido1990Aug,
	author = {Garrido, Pedro L. and Lebowitz, Joel L. and Maes, Christian and Spohn, Herbert},
	title = {{Long-range correlations for conservative dynamics}},
	journal = {Phys. Rev. A},
	volume = {42},
	number = {4},
	pages = {1954--1968},
	year = {1990},
	month = aug,
	publisher = {American Physical Society},
	doi = {10.1103/PhysRevA.42.1954}
}

@article{Maes1990Nov,
	author = {Maes, Christian},
	title = {{Kinetic limit of a conservative lattice gas dynamics showing long-range correlations}},
	journal = {J. Stat. Phys.},
	volume = {61},
	number = {3},
	pages = {667--681},
	year = {1990},
	month = nov,
	issn = {1572-9613},
	publisher = {Kluwer Academic Publishers-Plenum Publishers},
	doi = {10.1007/BF01027296}
}

@article{Rajesh2000May,
	author = {Rajesh, R. and Majumdar, Satya N.},
	title = {{Conserved Mass Models and Particle Systems in One Dimension}},
	journal = {J. Stat. Phys.},
	volume = {99},
	number = {3},
	pages = {943--965},
	year = {2000},
	month = may,
	issn = {1572-9613},
	publisher = {Kluwer Academic Publishers-Plenum Publishers},
	doi = {10.1023/A:1018651714376}
}

@article{Bondyopadhyay2012Jul,
	author = {Bondyopadhyay, Sourish and Mohanty, P. K.},
	title = {{Conserved mass models with stickiness and chipping}},
	journal = {J. Stat. Mech.: Theory Exp.},
	volume = {2012},
	number = {07},
	pages = {P07019},
	year = {2012},
	month = jul,
	issn = {1742-5468},
	publisher = {IOP Publishing and SISSA},
	doi = {10.1088/1742-5468/2012/07/P07019}
}

@article{Redig2017Oct,
	author = {Redig, Frank and Sau, Federico},
	title = {{Generalized immediate exchange models and their symmetries}},
	journal = {Stochastic Processes Appl.},
	volume = {127},
	number = {10},
	pages = {3251--3267},
	year = {2017},
	month = oct,
	issn = {0304-4149},
	publisher = {North-Holland},
	doi = {10.1016/j.spa.2017.02.005}
}

@article{Coppersmith1996May,
	author = {Coppersmith, S. N. and Liu, C.-h. and Majumdar, S. and Narayan, O. and Witten, T. A.},
	title = {{Model for force fluctuations in bead packs}},
	journal = {Phys. Rev. E},
	volume = {53},
	number = {5},
	pages = {4673--4685},
	year = {1996},
	month = may,
	issn = {2470-0053},
	publisher = {American Physical Society},
	doi = {10.1103/PhysRevE.53.4673}
}

@article{Lei2019Jan,
	author = {Qun-Li Lei and Massimo Pica Ciamarra and Ran Ni},
	title = {{Nonequilibrium strongly hyperuniform fluids of circle active particles with large local density fluctuations}},
	journal = {Science Advances},
	volume = {5},
	number = {1},
	year = {2019},
	month = jan,
	publisher = {American Association for the Advancement of Science},
	doi = {10.1126/sciadv.aau7423}
}

@article{VegaReyes2022Oct,
	author = {Vega Reyes, Francisco and L{\ifmmode\acute{o}\else\'{o}\fi}pez-Casta{\ifmmode\tilde{n}\else\~{n}\fi}o, Miguel A. and Rodr{\ifmmode\acute{\imath}\else\'{\i}\fi}guez-Rivas, {\ifmmode\acute{A}\else\'{A}\fi}lvaro},
	title = {{Diffusive regimes in a two-dimensional chiral fluid}},
	journal = {Commun. Phys.},
	volume = {5},
	number = {256},
	pages = {1--7},
	year = {2022},
	month = oct,
	issn = {2399-3650},
	publisher = {Nature Publishing Group},
	doi = {10.1038/s42005-022-01032-9}
}

@article{Green1954Mar,
	author = {Green, Melville S.},
	title = {{Markoff Random Processes and the Statistical Mechanics of Time‐Dependent Phenomena. II. Irreversible Processes in Fluids}},
	journal = {The Journal of Chemical Physics},
	volume = {22},
	number = {3},
	pages = {398--413},
	year = {1954},
	month = mar,
	publisher = {AIP Publishing},
	doi = {10.1063/1.1740082}
}

@article{Soni2019Nov,
	author = {Soni, Vishal and Bililign, Ephraim S. and Magkiriadou, Sofia and Sacanna, Stefano and Bartolo, Denis and Shelley, Michael J. and Irvine, William T. M.},
	title = {{The odd free surface flows of a colloidal chiral fluid}},
	journal = {Nat. Phys.},
	volume = {15},
	pages = {1188--1194},
	year = {2019},
	month = nov,
	issn = {1745-2481},
	publisher = {Nature Publishing Group},
	doi = {10.1038/s41567-019-0603-8}
}

@article{Yashunsky2022Nov,
	author = {Yashunsky, V. and Pearce, D. J. G. and Blanch-Mercader, C. and Ascione, F. and Silberzan, P. and Giomi, L.},
	title = {{Chiral Edge Current in Nematic Cell Monolayers}},
	journal = {Phys. Rev. X},
	volume = {12},
	number = {4},
	pages = {041017},
	year = {2022},
	month = nov,
	publisher = {American Physical Society},
	doi = {10.1103/PhysRevX.12.041017}
}

@article{Kubo1957Jun,
	author = {Kubo, Ryogo},
	title = {{Statistical-Mechanical Theory of Irreversible Processes. I. General Theory and Simple Applications to Magnetic and Conduction Problems}},
	journal = {J. Phys. Soc. Jpn.},
	volume = {12},
	number = {6},
	pages = {570--586},
	year = {1957},
	month = jun,
	issn = {0031-9015},
	publisher = {The Physical Society of Japan},
	doi = {10.1143/JPSJ.12.570}
}

@article{Avron1998Aug,
	author = {Avron, J. E.},
	title = {{Odd Viscosity}},
	journal = {Journal of Statistical Physics},
	volume = {92},
	number = {3},
	pages = {543--557},
	year = {1998},
	month = aug,
	publisher = {Kluwer Academic Publishers-Plenum Publishers},
	doi = {10.1023/A:1023084404080}
}

@article{Hargus2021Oct,
	author = {Cory Hargus and Jeffrey M. Epstein and Kranthi K. Mandadapu},
	title = {{Odd Diffusivity of Chiral Random Motion}},
	journal = {Physical Review Letters},
	volume = {127},
	number = {17},
	pages = {178001},
	year = {2021},
	month = oct,
	publisher = {American Physical Society},
	doi = {10.1103/PhysRevLett.127.178001}
}

@article{Hargus2025Mar,
	author = {Cory Hargus and Alhad Deshpande and Ahmad K. Omar and Kranthi K. Mandadapu},
	title = {{Flux Hypothesis for Odd Transport Phenomena}},
	journal = {Physical Review Letters},
	volume = {134},
	number = {9},
	pages = {097105},
	year = {2025},
	month = mar,
	publisher = {American Physical Society},
	doi = {10.1103/PhysRevLett.134.097105}
}

@article{Kalz2022Aug,
	author = {Erik Kalz and Hidde Derk Vuijk and Iman Abdoli and Jens-Uwe Sommer and Hartmut Löwen and Abhinav Sharma},
	title = {{Collisions Enhance Self-Diffusion in Odd-Diffusive Systems}},
	journal = {Physical Review Letters},
	volume = {129},
	number = {9},
	pages = {090601},
	year = {2022},
	month = aug,
	publisher = {American Physical Society},
	doi = {10.1103/PhysRevLett.129.090601}
}

@article{Chun2021Dec,
	author = {Hyun-Myung Chun and Qi Gao and Jordan M. Horowitz},
	title = {{Nonequilibrium Green-Kubo relations for hydrodynamic transport from an equilibrium-like fluctuation-response equality}},
	journal = {Physical Review Research},
	volume = {3},
	number = {4},
	pages = {043172},
	year = {2021},
	month = dec,
	publisher = {American Physical Society},
	doi = {10.1103/PhysRevResearch.3.043172}
}

@article{Cengio2019Dec,
	author = {Sara Dal Cengio and Demian Levis and Ignacio Pagonabarraga},
	title = {{Linear Response Theory and Green-Kubo Relations for Active Matter}},
	journal = {Physical Review Letters},
	volume = {123},
	number = {23},
	pages = {238003},
	year = {2019},
	month = dec,
	publisher = {American Physical Society},
	doi = {10.1103/PhysRevLett.123.238003}
}

@article{Berry1982Aug,
	author = {M V Berry},
	title = {{Wavelength-independent fringe spacing in rainbows from falling neutrons}},
	journal = {Journal of Physics A: Mathematical and General},
	volume = {15},
	number = {8},
	pages = {L385},
	year = {1982},
	month = aug,
	publisher = {IOP Publishing},
	doi = {10.1088/0305-4470/15/8/001}
}

@article{Kravtsov1983Dec,
	author = {Yu A Kravtsov and Yu I Orlov},
	title = {{Caustics, catastrophes, and wave fields}},
	journal = {Soviet Physics Uspekhi},
	volume = {26},
	number = {12},
	pages = {1038},
	year = {1983},
	month = dec,
	publisher = {IOP Publishing},
	doi = {10.1070/PU1983v026n12ABEH004582}
}

@incollection{berry1980catastrophe,
  title     = {IV Catastrophe Optics: Morphologies of Caustics and Their Diffraction Patterns},
  author    = {Berry, M. V. and Upstill, C.},
  booktitle = {Progress in Optics},
  volume    = {18},
  pages     = {257--346},
  year      = {1980},
  publisher = {Elsevier},
  doi       = {10.1016/S0079-6638(08)70215-4}
}

@article{Dolai2025Oct,
	author = {Dolai, Pritha and Das, Arghya},
	title = {{Boundary layers, transport and universal distribution in boundary driven active systems}},
	journal = {SciPost Physics},
	volume = {19},
	number = {4},
	pages = {088},
	year = {2025},
	month = oct,
	doi = {10.21468/SciPostPhys.19.4.088}
}

@article{Johnson1928Jul,
	author = {J. B. Johnson},
	title = {{Thermal Agitation of Electricity in Conductors}},
	journal = {Physical Review},
	volume = {32},
	number = {1},
	pages = {97},
	year = {1928},
	month = jul,
	publisher = {American Physical Society},
	doi = {10.1103/PhysRev.32.97}
}

@article{Ginot2025Nov,
	author = {Ginot, Félix and Bechinger, Clemens},
	title = {{Energy recuperation of driven colloids in non-Markovian baths}},
	journal = {Nature Communications},
	volume = {16},
	number = {1},
	pages = {10114},
	year = {2025},
	month = nov,
	publisher = {Nature Publishing Group},
	doi = {10.1038/s41467-025-66191-z}
}

@article{Marchetti2013Jul,
	author = {M. C. Marchetti and J. F. Joanny and S. Ramaswamy and T. B. Liverpool and J. Prost and Madan Rao and R. Aditi Simha},
	title = {{Hydrodynamics of soft active matter}},
	journal = {Reviews of Modern Physics},
	volume = {85},
	number = {3},
	pages = {1143},
	year = {2013},
	month = jul,
	publisher = {American Physical Society},
	doi = {10.1103/RevModPhys.85.1143}
}

@article{Doyle1998Apr,
	author = {Declan A. Doyle and João Morais Cabral and Richard A. Pfuetzner and Anling Kuo and Jacqueline M. Gulbis and Steven L. Cohen and Brian T. Chait and Roderick MacKinnon},
	title = {{The Structure of the Potassium Channel: Molecular Basis of K+ Conduction and Selectivity}},
	journal = {Science},
	volume = {280},
    number={5360},
	pages = {69-77},
	year = {1998},
	month = apr,
	publisher = {American Association for the Advancement of Science},
	doi = {10.1126/science.280.5360.69}
}

@article{Maes2020Mar,
    author = {Christian Maes},
	title = {{Frenesy: Time-symmetric dynamical activity in nonequilibria}},
	journal = {Physics Reports},
	volume = {850},
	pages = {1--33},
	year = {2020},
	month = mar,
	publisher = {North-Holland},
	doi = {10.1016/j.physrep.2020.01.002}
}

@article{Dutta1981Jul,
	author = {P. Dutta and P. M. Horn},
	title = {{Low-frequency fluctuations in solids: <span class="aps-inline-formula"><math xmlns="http://www.w3.org/1998/Math/MathML" display="inline"><mfrac><mrow><mn>1</mn></mrow><mrow><mi>f</mi></mrow></mfrac></math></span> noise}},
	journal = {Reviews of Modern Physics},
	volume = {53},
	number = {3},
	pages = {497},
	year = {1981},
	month = jul,
	publisher = {American Physical Society},
	doi = {10.1103/RevModPhys.53.497}
}

@book{Leal2007Jun,
	author = {L. Gary Leal},
	title = {{Advanced Transport Phenomena: Fluid Mechanics and Convective Transport Processes}},
	publisher = {Cambridge University Press},
	year = {2007},
	month = jun,
	doi = {10.1017/CBO9780511800245}
}

@article{Gelbart1996Aug,
	author = {Gelbart, William M. and Ben-Shaul, Avinoam},
	title = {{The “New” Science of “Complex Fluids”}},
	journal = {The Journal of Physical Chemistry},
	volume = {100},
	number = {31},
	pages = {13169--13189},
	year = {1996},
	month = aug,
	publisher = {American Chemical Society},
	doi = {10.1021/jp9606570}
}

@article{Fruchart2023Mar,
	author = {Michel Fruchart and Colin Scheibner and Vincenzo Vitelli},
	title = {{Odd Viscosity and Odd Elasticity}},
	journal = {Annual Review of Condensed Matter Physics},
	volume = {14},
	number = {Volume 14, 2023},
	pages = {471--510},
	year = {2023},
	month = mar,
	publisher = {Annual Reviews},
	doi = {10.1146/annurev-conmatphys-040821-125506}
}

@article{Nguyen2014Feb,
	author = {Nguyen H. P. Nguyen and Daphne Klotsa and Michael Engel and Sharon C. Glotzer},
	title = {{Emergent Collective Phenomena in a Mixture of Hard Shapes through Active Rotation}},
	journal = {Physical Review Letters},
	volume = {112},
	number = {7},
	pages = {075701},
	year = {2014},
	month = feb,
	publisher = {American Physical Society},
	doi = {10.1103/PhysRevLett.112.075701}
}

@article{Guo2025Jan,
	author = {Rui-xue Guo and Jia-jian Li and Bao-quan Ai},
	title = {{Diffusion of active particles driven by odd interactions}},
	journal = {Physical Review E},
	volume = {111},
	number = {1},
	pages = {014105},
	year = {2025},
	month = jan,
	publisher = {American Physical Society},
	doi = {10.1103/PhysRevE.111.014105}
}

@article{Kummel2013May,
	author = {Felix Kümmel and Borge ten Hagen and Raphael Wittkowski and Ivo Buttinoni and Ralf Eichhorn and Giovanni Volpe and Hartmut Löwen and Clemens Bechinger},
	title = {{Circular Motion of Asymmetric Self-Propelling Particles}},
	journal = {Physical Review Letters},
	volume = {110},
	number = {19},
	pages = {198302},
	year = {2013},
	month = may,
	publisher = {American Physical Society},
	doi = {10.1103/PhysRevLett.110.198302}
}

@article{Hasan2010Nov,
	author = {M. Z. Hasan and C. L. Kane},
	title = {{<i>Colloquium</i>: Topological insulators}},
	journal = {Reviews of Modern Physics},
	volume = {82},
	number = {4},
	pages = {3045},
	year = {2010},
	month = nov,
	publisher = {American Physical Society},
	doi = {10.1103/RevModPhys.82.3045}
}

@article{Avron1995Jul,
	author = {J. E. Avron and R. Seiler and P. G. Zograf},
	title = {{Viscosity of Quantum Hall Fluids}},
	journal = {Physical Review Letters},
	volume = {75},
	number = {4},
	pages = {697},
	year = {1995},
	month = jul,
	publisher = {American Physical Society},
	doi = {10.1103/PhysRevLett.75.697}
}

@article{Miyake2013Oct,
	author = {Hirokazu Miyake and Georgios A. Siviloglou and Colin J. Kennedy and William Cody Burton and Wolfgang Ketterle},
	title = {{Realizing the Harper Hamiltonian with Laser-Assisted Tunneling in Optical Lattices}},
	journal = {Physical Review Letters},
	volume = {111},
	number = {18},
	pages = {185302},
	year = {2013},
	month = oct,
	publisher = {American Physical Society},
	doi = {10.1103/PhysRevLett.111.185302}
}

@article{Hugel2014Feb,
	author = {Dario Hügel and Belén Paredes},
	title = {{Chiral ladders and the edges of quantum Hall insulators}},
	journal = {Physical Review A},
	volume = {89},
	number = {2},
	pages = {023619},
	year = {2014},
	month = feb,
	publisher = {American Physical Society},
	doi = {10.1103/PhysRevA.89.023619}
}

@article{Banerjee2022Apr,
	author = {Debarghya Banerjee and Anton Souslov and Vincenzo Vitelli},
	title = {{Hydrodynamic correlation functions of chiral active fluids}},
	journal = {Physical Review Fluids},
	volume = {7},
	number = {4},
	pages = {043301},
	year = {2022},
	month = apr,
	publisher = {American Physical Society},
	doi = {10.1103/PhysRevFluids.7.043301}
}

@article{Liebchen2017Aug,
	author = {Benno Liebchen and Demian Levis},
	title = {{Collective Behavior of Chiral Active Matter: Pattern Formation and Enhanced Flocking}},
	journal = {Physical Review Letters},
	volume = {119},
	number = {5},
	pages = {058002},
	year = {2017},
	month = aug,
	publisher = {American Physical Society},
	doi = {10.1103/PhysRevLett.119.058002}
}

@article{Liebchen2022Sep,
	author = {Benno Liebchen and Demian Levis},
	title = {{Chiral active matter}},
	journal = {Europhysics Letters},
	volume = {139},
	number = {6},
	pages = {67001},
	year = {2022},
	month = sep,
	publisher = {IOP Publishing},
	doi = {10.1209/0295-5075/ac8f69}
}

@article{Eren2026Aug,
	author = {Ege Eren and Michel Fruchart and Vincenzo Vitelli},
	title = {{Hydrodynamic description of collisional odd fluids from kinetic theory}},
	journal = {Physical Review E},
	volume = {114},
	number = {2},
	pages = {024109},
	year = {2026},
	month = aug,
	publisher = {American Physical Society},
	doi = {10.1103/v57s-653w}
}

@article{vanZuiden2016Nov,
	author = {van Zuiden, Benjamin C. and Paulose, Jayson and Irvine, William T. M. and Bartolo, Denis and Vitelli, Vincenzo},
	title = {{Spatiotemporal order and emergent edge currents in active spinner materials}},
	journal = {Proceedings of the National Academy of Sciences},
	volume = {113},
	number = {46},
	pages = {12919--12924},
	year = {2016},
	month = nov,
	publisher = {Proceedings of the National Academy of Sciences},
	doi = {10.1073/pnas.1609572113}
}

@article{Levis2019Jul,
	author = {Demian Levis and Benno Liebchen},
	title = {{Simultaneous phase separation and pattern formation in chiral active mixtures}},
	journal = {Physical Review E},
	volume = {100},
	number = {1},
	pages = {012406},
	year = {2019},
	month = jul,
	publisher = {American Physical Society},
	doi = {10.1103/PhysRevE.100.012406}
}

@article{Caporusso2024Apr,
	author = {Claudio B. Caporusso and Giuseppe Gonnella and Demian Levis},
	title = {{Phase Coexistence and Edge Currents in the Chiral Lennard-Jones Fluid}},
	journal = {Physical Review Letters},
	volume = {132},
	number = {16},
	pages = {168201},
	year = {2024},
	month = apr,
	publisher = {American Physical Society},
	doi = {10.1103/PhysRevLett.132.168201}
}

@article{Caprini2025Apr,
	author = {Caprini, Lorenzo and Marini Bettolo Marconi, U.},
	title = {{Bubble phase induced by odd interactions in chiral systems}},
	journal = {The Journal of Chemical Physics},
	volume = {162},
	number = {16},
	year = {2025},
	month = apr,
	publisher = {AIP Publishing},
	doi = {10.1063/5.0262594}
}

@article{Kalz2024Jan,
	author = {Erik Kalz and Hidde Derk Vuijk and Jens-Uwe Sommer and Ralf Metzler and Abhinav Sharma},
	title = {{Oscillatory Force Autocorrelations in Equilibrium Odd-Diffusive Systems}},
	journal = {Physical Review Letters},
	volume = {132},
	number = {5},
	pages = {057102},
	year = {2024},
	month = jan,
	publisher = {American Physical Society},
	doi = {10.1103/PhysRevLett.132.057102}
}

@article{Lowen2016Nov,
	author = {Löwen, Hartmut},
	title = {{Chirality in microswimmer motion: From circle swimmers to active turbulence}},
	journal = {The European Physical Journal Special Topics},
	volume = {225},
	number = {11},
	pages = {2319--2331},
	year = {2016},
	month = nov,
	publisher = {Springer Berlin Heidelberg},
	doi = {10.1140/epjst/e2016-60054-6}
}

@article{Shankar2022Jun,
	author = {Shankar, Suraj and Souslov, Anton and Bowick, Mark J. and Marchetti, M. Cristina and Vitelli, Vincenzo},
	title = {{Topological active matter}},
	journal = {Nature Reviews Physics},
	volume = {4},
	number = {6},
	pages = {380--398},
	year = {2022},
	month = jun,
	publisher = {Nature Publishing Group},
	doi = {10.1038/s42254-022-00445-3}
}

@article{Larralde1997Nov,
	author = {H. Larralde},
	title = {{Transport properties of a two-dimensional “chiral” persistent random walk}},
	journal = {Physical Review E},
	volume = {56},
	number = {5},
	pages = {5004},
	year = {1997},
	month = nov,
	publisher = {American Physical Society},
	doi = {10.1103/PhysRevE.56.5004}
}

@article{Marconi2008March,
title = {Fluctuation–dissipation: Response theory in statistical physics},
journal = {Physics Reports},
volume = {461},
number = {4},
pages = {111-195},
year = {2008},
issn = {0370-1573},
doi = {https://doi.org/10.1016/j.physrep.2008.02.002},
url = {https://www.sciencedirect.com/science/article/pii/S0370157308000768},
author = {Umberto Marini Bettolo Marconi and Andrea Puglisi and Lamberto Rondoni and Angelo Vulpiani},
}

@article{Speck2006May,
	author = {T. Speck and U. Seifert},
	title = {{Restoring a fluctuation-dissipation theorem in a nonequilibrium steady state}},
	journal = {Europhysics Letters},
	volume = {74},
	number = {3},
	pages = {391--396},
	year = {2006},
	month = may,
	publisher = {EDP Sciences},
	doi = {10.1209/epl/i2005-10549-4}
}

@article{Baiesi2009Jul,
	author = {Marco Baiesi and Christian Maes and Bram Wynants},
	title = {{Fluctuations and Response of Nonequilibrium States}},
	journal = {Physical Review Letters},
	volume = {103},
	number = {1},
	pages = {010602},
	year = {2009},
	month = jul,
	publisher = {American Physical Society},
	doi = {10.1103/PhysRevLett.103.010602}
}

@article{Blickle2007May,
	author = {V. Blickle and T. Speck and C. Lutz and U. Seifert and C. Bechinger},
	title = {{Einstein Relation Generalized to Nonequilibrium}},
	journal = {Physical Review Letters},
	volume = {98},
	number = {21},
	pages = {210601},
	year = {2007},
	month = may,
	publisher = {American Physical Society},
	doi = {10.1103/PhysRevLett.98.210601}
}

@article{Einstein1905Jan,
	author = {A. Einstein},
	title = {{Über die von der molekularkinetischen Theorie der Wärme geforderte Bewegung von in ruhenden Flüssigkeiten suspendierten Teilchen}},
	journal = {Annalen der Physik},
	volume = {322},
	number = {8},
	pages = {549--560},
	year = {1905},
	month = jan,
	publisher = {John Wiley \& Sons, Ltd},
	doi = {10.1002/andp.19053220806}
}

@article{Nyquist1928Jul,
	author = {H. Nyquist},
	title = {{Thermal Agitation of Electric Charge in Conductors}},
	journal = {Physical Review},
	volume = {32},
	number = {1},
	pages = {110},
	year = {1928},
	month = jul,
	publisher = {American Physical Society},
	doi = {10.1103/PhysRev.32.110}
}

@article{Wang2024SepNJP,
	author = {Boyi Wang and Frank Jülicher and Patrick Pietzonka},
	title = {{Condensate formation in a chiral lattice gas}},
	journal = {New Journal of Physics},
	volume = {26},
	number = {9},
	pages = {093031},
	year = {2024},
	month = sep,
	publisher = {IOP Publishing},
	doi = {10.1088/1367-2630/ad7490}
}

@article{Dutta2026Sep,
	author = {Aditya Kumar Dutta and Swarnajit Chatterjee and Matthieu Mangeat and Raja Paul},
	title = {{Stability and breakdown of chiral motion in nonreciprocal flocking}},
	journal = {Physical Review E},
	volume = {114},
	number = {3},
	pages = {034115},
	year = {2026},
	month = sep,
	publisher = {American Physical Society},
	doi = {10.1103/yz2d-wk6g}
}

@article{Sevilla2016Dec,
	author = {Sevilla, Francisco J.},
	title = {{Diffusion of active chiral particles}},
	journal = {Physical Review E},
	volume = {94},
	number = {6},
	year = {2016},
	month = dec,
	publisher = {American Physical Society (APS)},
	doi = {10.1103/PhysRevE.94.062120}
}

@article{Caprini2019Mar,
	author = {Caprini, Lorenzo and Marini Bettolo Marconi, Umberto},
	title = {{Active chiral particles under confinement: surface currents and bulk accumulation phenomena}},
	journal = {Soft Matter},
	volume = {15},
	number = {12},
	pages = {2627--2637},
	year = {2019},
	month = mar,
	publisher = {The Royal Society of Chemistry},
	doi = {10.1039/c8sm02492h}
}

\end{document}